\documentclass[]{elsarticle}

\journal{International Journal of Heat and Fluid Flow}

\usepackage{numcompress}\biboptions{authoryear}
\usepackage[pdftex]{}

\usepackage{amsmath}

\newcommand{\pd}[2]{\frac{\partial #1}{\partial #2}}
\newcommand{\pda}[2]{\frac{\partial #1 ^\ast}{\partial #2 ^\ast}}

\newcommand{\Rew}{\textrm{Re}_{\rm w}}

\newcommand{\Wew}{\textrm{Wi}_{\rm w}}

\begin{document}

\begin{frontmatter}

\title{Viscoelasticity-induced pulsatile motion of 2D roll cell in laminar wall-bounded shear flow}

 \author{Tomohiro Nimura}
 \author{Takuya Kawata}
 \author{Takahiro Tsukahara\corref{cor1}}
 \ead{tsuka@rs.tus.ac.jp}
 \cortext[cor1]{Corresponding author:}

\address{Department of Mechanical Engineering, Tokyo University of Science, 2641 Yamazaki, Noda-shi, Chiba 278-8510, Japan}

\begin{abstract}
For the clarification of the routes to elasto-inertial turbulence (EIT), it is essential to understand how viscoelasticity modulates coherent flow structures including the longitudinal vortices. 
We focused on a rotating plane Couette flow that provides two-dimensional (2D) roll cells for the steady laminar Newtonian-fluid case, and we investigated how the steady longitudinal vortices are modulated by viscoelasticity at different Weissenberg numbers. 
The viscoelasticity was found to induce an unsteady flow state where the 2D roll-cell structure was periodically enhanced and damped with a constant period, keeping the homogeneity in the streamwise direction. 
This pulsatile motion of the roll cell was caused by a time lag in the response of the viscoelastic force to the vortex development. 
Both the pulsation period and time lag were found to be scaled by the turnover time of cell rotation rather than by the relaxation time, despite the viscoelasticity-induced instability.
We also discuss the counter torque on the roll cell and the net energy balance, considering their relevance to polymer drag reduction and EIT.
\end{abstract}

\begin{keyword}
DNS \sep drag reduction \sep rotating plane Couette flow \sep viscoelastic fluid \sep Giesekus model \sep wall turbulence
\end{keyword}

\end{frontmatter}


\section{Introduction}
\label{intro}
Viscoelasticity not only causes turbulent drag reduction (DR), but also enhances disordered motions at higher additive concentration. 
In drag-reducing wall turbulence, the elastic force due to the additive tends to suppress the secondary motions of the near-wall coherent structures and, therefore, the near-wall structures are modulated to be apparently streamwise-independent streaks \citep{sureshkumar97,stone02,stone04,kim07}. 
The other aspect of the viscoelastic effect that yields an instability at high additive concentrations has also been subject of many studies since Giesekus's discovery of elasticity-induced instabilities \citep{giesekus72}. 
\cite{larson90} predicted that viscoelasticity gives rise to an oscillating mode by analyzing the linear stability of an inertia-less flow of Oldroyd-B fluid. 
Moreover, in flows at the state of maximum drag reduction (MDR), the viscoelasticity promotes a transition to chaotic flow state even at very low Reynolds numbers, which is known as elasto-inertial turbulence (EIT) \citep{Hoyt77,groisman98,Dubief13,Samanta13,Pan13,terrapon15}.

Recently, the possible connection between EIT and MDR has been pointed out. 
Once DR occurs, weakened streamwise vortices disappear temporarily for a certain period, during which the velocity profile approaches the MDR asymptote \citep{virk75}. 
Such temporal behavior, called hibernation, is known to occur even in the Newtonian wall turbulence; however, its frequency increases in the viscoelastic drag-reducing turbulence~\citep{xi10}. 
The hibernating turbulence and the aforementioned viscoelastic effect to modulate the near-wall structures are commonly observed in both MDR and EIT flows~\citep{xi10,Samanta13,Dubief13,terrapon15}. 
Recent studies have suggested the MDR phenomenon as a part of the EIT at the significant large Weissenberg number. 
\cite{sid18} demonstrated through numerical simulation that at high enough additive concentration a two-dimensional and chaotic flow can be sustained after sufficient perturbations have been introduced. This flow state was characterized by an energy injection from the additive to flow at medium and small scales. 
According to another work by \cite{choueiri18}, such two-dimensional structures would appear when the additive concentration exceeds the MDR asymptote, whereas in the regime before the MDR asymptote the streamwise vortices are still observed with hibernating behaviors. 
These observations imply that the influence of viscoelasticity would be dominant in the EIT and change its role from that for the MDR. 
Because of the complexity of background turbulence, the route to EIT is still not well understood, in particular, regarding the transition process from the longitudinal vortices in DR to other forms that can be observed only in MDR or EIT.
In this context, we can consider the elasticity-induced modulation of the coherent structures as a key phenomenon to understand the transition mechanism from DR to MDR and EIT.

\begin{figure}[t]
\begin{center}
	\includegraphics[width=0.66\hsize]{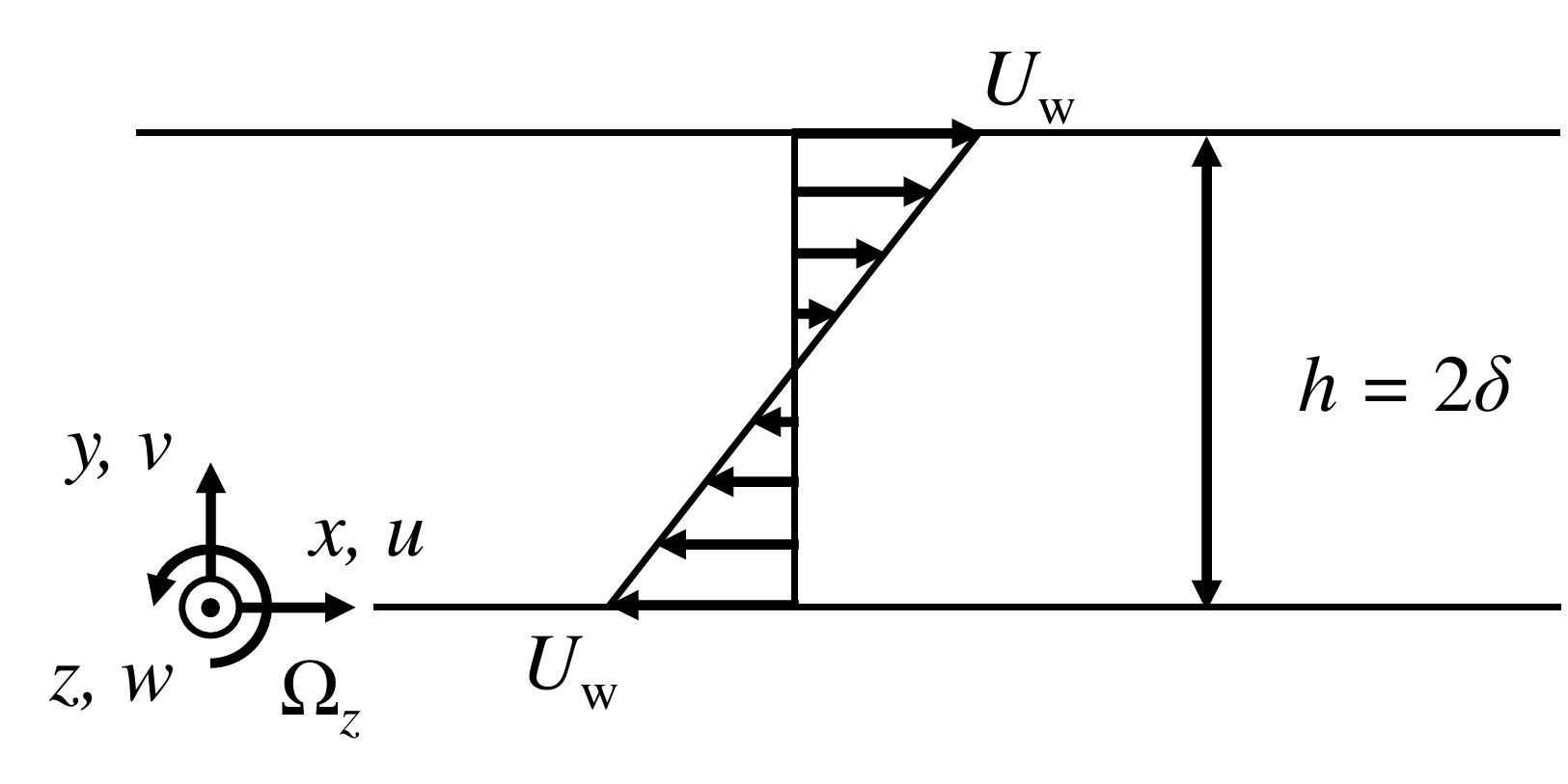}
	\caption{Configuration of the rotating plane Couette flow.}
	\label{fig:RPCF}
\end{center}
\end{figure}

Now, let us introduce the plane Couette flow subjected to spanwise system rotation (rotating plane Couette flow, RPCF), the definition of which is schematically shown in Fig.~\ref{fig:RPCF}. 
This flow is known to bring distinct coherent streamwise roll cells and can therefore be a good test bench to study the effect of viscoelasticity on longitudinal vortices in a wall-bounded shear flow. 
Given an anti-cyclonic system rotation, where the system rotates in the opposite direction to the wall shear, the flow is linearly unstable owing to the Coriolis-force effect, which gives rise to the streamwise-elongated roll-cell structure. 
Depending on the Reynolds number $\Rew$ and the rotation number $\Omega$ (the definitions are given in the next section), the coherent roll cells can take various forms, such as two-dimensional (2D) steady roll cells and three-dimensional (3D) wavy roll cells \citep{tsuka10, Kawata16a, kawata16b}. 
Comparing the flow structures in the RPCFs of Newtonian and viscoelastic fluids, one may gain physical insights into how the viscoelasticity modulates the longitudinal vortices in shear flow and why it leads to DR or EIT. 

In this work, we performed direct numerical simulations (DNSs) to study the laminar RPCF of viscoelastic fluid at a Reynolds number of $\Rew=25$ and a rotation number of $\Omega=10$ (the definitions of these parameters are given in the next section) over a wide range of Weissenberg numbers. 
This set of control-parameter values is chosen as a typical flow case that gives a steady and streamwise-independent roll cells in the Newtonian case, to better understand how the increase in viscoelasticity affects the instabilities of a streamwise vortical structure. We show that the increase in the viscoelasticity effect gives rise to an unsteady flow state where the 2D roll cells are periodically strengthened and suppressed, 
the time scale of which is on the same order as the hibernation period in the drag-reduced turbulence that was found by \cite{xi10}. 
We also discuss the energy exchange between the flow and the additive, and show that, in the pulsatile flow state, there is a certain time lag between the change in the flow structure and the energy exchange. 
Through these analyses, we demonstrate a negative torque on the roll cell in relation to the DR as well as the onset of unsteadiness that could be linked with EIT.

\section{Numerical method}
\label{method}

\subsection{Flow configuration and governing equations}

The coordinate system is defined as shown in Fig.~\ref{fig:RPCF}: the $x$-, $y$-, and $z$-axes are taken in the streamwise, wall-normal, and spanwise directions, respectively. 
The top and bottom walls are located at $y=h$ and $y=0$, respectively, and they move in opposite directions with a speed of $U_{\rm w}$. The Reynolds number $\Rew$ and the rotation number $\Omega$ are defined, on the basis of the wall speed $U_{\rm w}$ and the half channel height $\delta$ $(=h/2)$, as $\Rew= U_{\rm w} \delta / \nu$ and $\Omega=2\Omega_z \delta^2 / \nu$, respectively, where $\nu$ is the kinematic viscosity at zero shear rate. 

The governing equations solved numerically in the present DNS are the nondimensional continuity and the non-Newtonian momentum equations written in a frame of reference rotating with the system: 
\begin{equation}
   \pda{u_i}{x_i} = 0, 
   \label{eq:cont}
\end{equation}
and
\begin{eqnarray}
\pda{u_i}{t}+u_j^\ast \pda{u_i}{x_j} &=& -\pda{p}{x_i} + \frac{\beta}{\Rew} \pd{^2 u_i^\ast}{{x_j^\ast}^2}  \nonumber \\ 
&&- \frac{\Omega}{\Rew} \epsilon_{i3k} u_k^\ast + \frac{1- \beta}{\Wew} \pd{c_{ij}}{x_j^\ast}.
\label{eq:NS}
\end{eqnarray}
Here, $p$ is the pressure hydrostatic including both the static pressure and the centrifugal acceleration, $\epsilon_{ijk}$ is the Levi--Civita symbol, and the variables with the superscript $\ast$ stand for the nondimensional quantities normalized by $\delta$ and/or $U_{\rm w}$. 
The viscosity ratio and the Weissenberg number are defined as $\beta = \mu_{\rm s} / (\mu_{\rm s} + \mu_{\rm a} )$ and $\Wew = U_{\rm w}^2 \lambda / \nu$, where $\mu_{\rm s}$ and $\mu_{\rm a}$ are the viscosity of the solution and the additive, respectively, and $\lambda$ is the relaxation time of the additive. 
The former $\beta$ is a measure of the concentration of the additive, and the effect of the viscoelasticity becomes more significant with decreasing $\beta$. 
The Weissenberg number $\Wew$ physically represents the ratio of the relaxation time of the additive to the viscous time scale. 
The nondimensional conformation tensor $c_{ij}$ of the last term in Eq.~(\ref{eq:NS}) is defined on the basis of the extra stress tensor by viscoelasticity $\tau_{ij}$ as $c_{ij}=\tau_{ij} \lambda / \mu_\mathrm{a} + \delta_{ij}$ (where $\delta_{ij}$ is the Kronecker delta) and is governed by a constitutive equation. 
We adopted the Giesekus model \citep{Giesekus82}: 
\begin{eqnarray}
\pd{c_{ij}}{t^\ast} + \pd{u_m^\ast  c_{ij}}{x_m^\ast} -\pd{u_i^\ast}{x_m^\ast} c_{mj} -\pd{u_j^\ast}{x_m^\ast} c_{mi} \hspace{4cm} \nonumber \\
+\frac{\Rew}{\Wew} \left[ c_{ij}  -\delta_{ij} + \alpha (c_{im} - \delta_{im})(c_{mj} - \delta_{mj}) \right] = 0, 
\label{eq:Giesekus}
\end{eqnarray}
where $\alpha$ is the mobility factor, the value of which is between 0 and 1. 
The mobility factor $\alpha$ represents the strength of the nonlinearity effect in the Giesekus model and is known to be proportional to the inverse of the maximum polymer extension in the finitely extensible nonlinear elastic-Peterlin (FENE-P) model. Hence, the elastic scales are smaller with increasing $\alpha$.

In the present study, the viscosity ratio and the mobility factor were fixed at $\beta=0.8$ and $\alpha=0.001$, respectively, as the DNS on a plane or an orifice channel flow performed with these values of $\alpha$ and $\beta$ \citep{tsuka11,tsuka13} showed a qualitative agreement in terms of the DR effect with the experiment result using a drag-reducing surfactant. 

\subsection{Numerical procedures}

We used the finite difference method for the spatial discretization. 
The fourth-order central difference scheme was used for the $x$- and $z$-directions, whereas the second-order central difference scheme was adopted in the wall-normal ($y$-) direction. 
For the time integration, the second-order Crank--Nicolson and the second-order Adams--Bashforth schemes were used for the wall-normal viscous term and the other terms, respectively. 
As for the constitutive equations with the Giesekus model, a flux limiter of the MINMOD scheme was adopted to approximate the spatial derivatives in the advective term without adding artificial diffusivity, as \cite{yu04} proposed. 
As for the boundary condition, the periodic boundary conditions were imposed in the $x$- and $z$- directions and the no-slip condition was applied on the walls.

In the present study, we employed a computational domain size of $L_x \times L_y \times L_z=7.5h \times 2\delta \times 2h$, to massively limit the degree of freedom artificially and thereby extract the essential influence of increasing $\Wew$ on the flow structure. 
The streamwise domain length $L_x=7.5h$ corresponded to the streamwise wavelength of the 3D wavy roll cells experimentally observed in the Newtonian RPCF \citep{tsuka10}, and the spanwise domain length $L_z=2h$ was even smaller than the spanwise scale of the experimentally observed structure. 
In a computation with a smaller domain size, the 2D roll cells in the Newtonian case did not develop. 
In the computation with a larger domain size, on the other hand, the observed tendency of the ${\rm Wi}$ effect remained qualitatively the same, whereas the viscoelastic instabilities observed in the flow structure were more complex. 
The grid number was 128 in all directions; however, the nonuniform grid was applied in the wall-normal direction: the grid sizes were comparable to or smaller than the viscous length. 
As previously mentioned, we focused only on the case of $\Rew=25$ and $\Omega=10$ in the present study, and the DNS results for several different Weissenberg numbers in the range $\Wew=0$--2000 are compared in the following sections.

Regarding the initial conditions, as the RPCF at $\Rew=25$ and $\Omega=10$ was linearly unstable, the unstable mode exponentially grew independently of the initial disturbance. In the present study, we gave velocity disturbances with a certain magnitude just to make the flow develop fully earlier, and random noise with a magnitude of 50\% of $U_\mathrm{w}$ was superimposed on the linear velocity profile of the laminar plane Couette flow as the initial condition of the Newtonian case, whereas in the viscoelastic cases a pair of streamwise-independent vortices with a maximum swirling velocity of 0.1\% of $U_\mathrm{w}$ located at the center of the computational domain was used. We also tested the robustness and grid convergence of the flow structure finally obtained in the present study by using different types of initial disturbances and different grid sizes, which is described in the Appendix.

\section{Results \& Discussion}
\label{result}
In the following, we denote the mean velocities spatially averaged in the $x$- and $z$-directions by $\overline{u_i}$, and $u_i^\prime$ represents the deviation from the mean values: $u_i^\prime=u_i-\overline{u_i}$. An expression for the time-averaged quantity is denoted by $\langle \cdot \rangle_t$. We use lowercase and uppercase characters to denote the local and the volume-averaged quantities, respectively. 

\subsection{Weissenberg-number dependency of growth rate}
\label{wei_depend}

\begin{figure}[t]
\begin{center}
	\includegraphics[width=0.75\hsize]{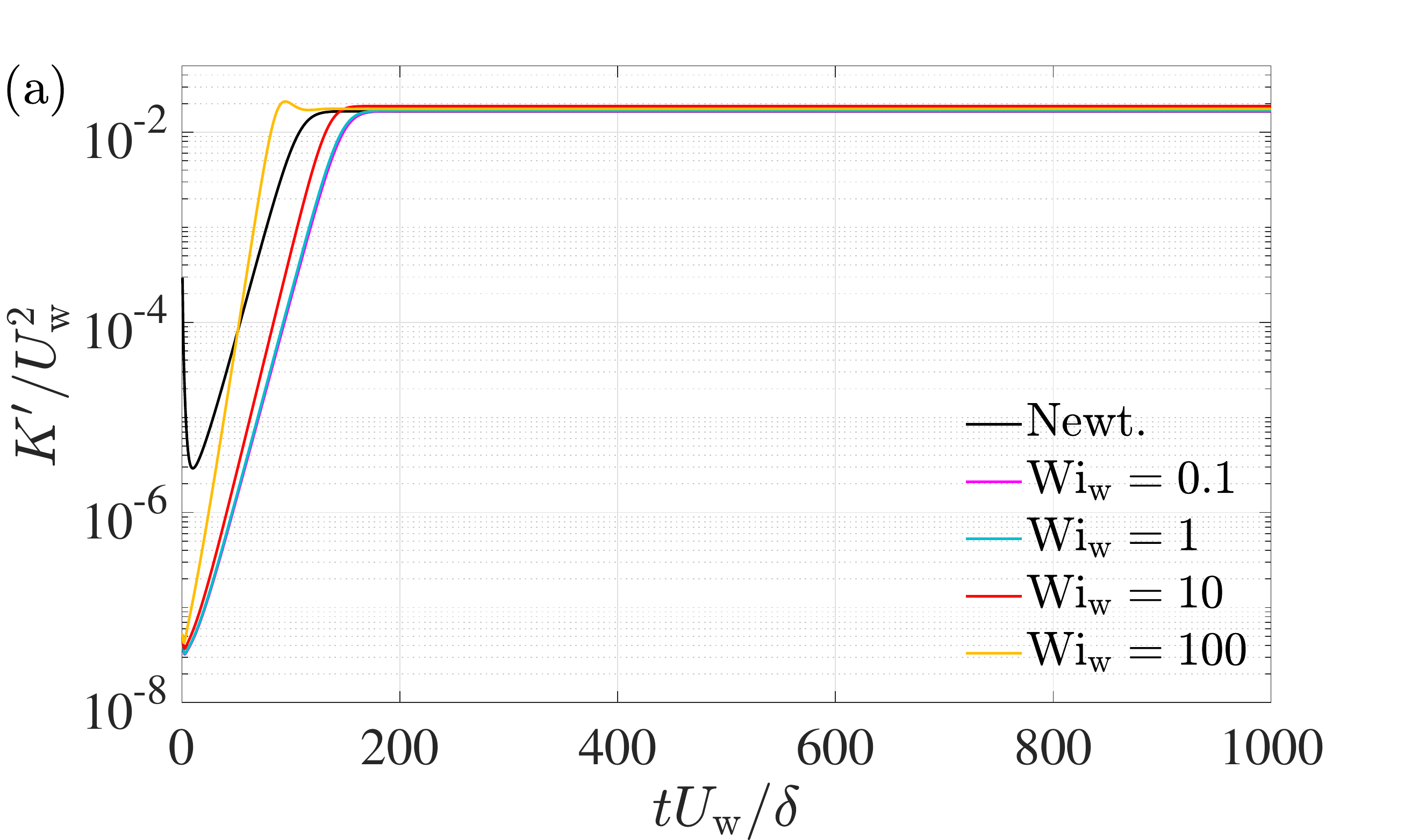}
	\includegraphics[width=0.75\hsize]{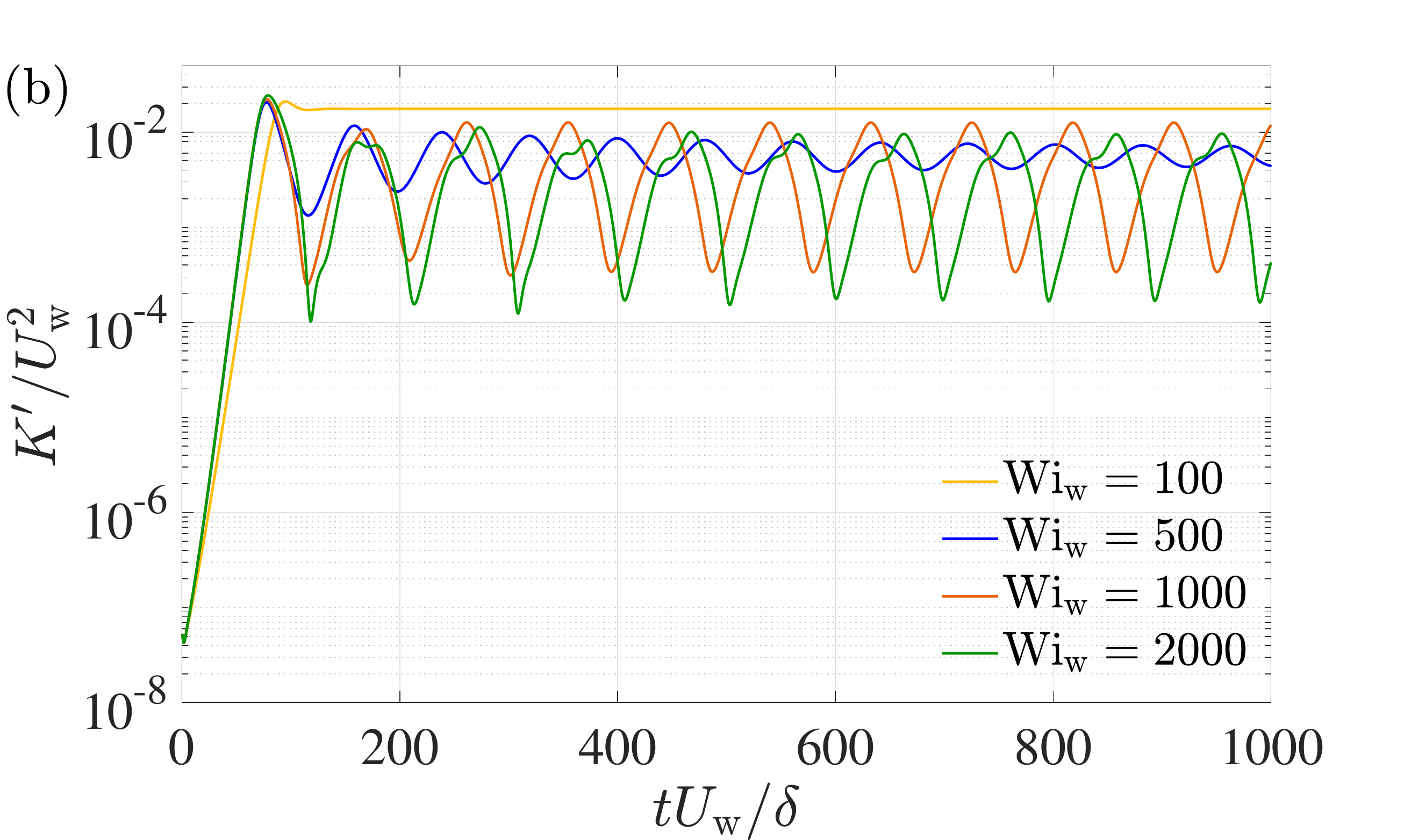}
	\caption{Time evolution of the volume-averaged kinetic energy $K^{\prime}$ of the secondary motion in Newtonian (Newt.) and various Weissenberg-number cases: (a) lower Weissenberg numbers and (b) higher ones. }
	\label{fig:K_t}
\end{center}
\end{figure}

Figure~\ref{fig:K_t} shows the time evolution of the volume-averaged kinetic energy defined on the basis of the velocity deviations from the mean value:
\begin{equation}
      K^{\prime}=\frac{1}{L_x L_y L_z}  \int^{L_z}_0 \int^{L_y}_0 \int^{L_x}_0 \frac{1}{2}({u^\prime}^2+{v^\prime}^2+{w^\prime}^2) {\rm d}x{\rm d}y{\rm d}z. 
      \label{eq:K}
\end{equation}
\noindent

As shown in Fig.~\ref{fig:K_t}(a), the initial condition in the Newtonian case had a relatively larger kinetic energy compared to the other viscoelastic cases, as already mentioned in the previous section; however, in all the Newtonian and viscoelastic cases, the time evolutions of $K^\prime$ showed exponential growths at the early stage of the simulations, approximately $0 < t^{\ast} <150$. 
In the Newtonian and viscoelastic cases with relatively low Weissenberg numbers $\Wew \leq 100$, the flow reached a steady state, and it is also shown that, in the case of $\Wew=100$, the growth of $K^\prime$ exhibited an overshoot before the steady state. 
For the higher Weissenberg-number cases shown in Fig.~\ref{fig:K_t}(b), the time evolutions of $K^\prime$ exhibited periodic behaviors. 
At $\Wew=500$, the amplitude of the periodic behavior decreased in time, and after a long time had elapsed, such a periodic unsteadiness disappeared. 
The mean value $\langle K^\prime \rangle_t$ after the initial growth phase of $K^\prime$ was significantly lower than that obtained at $\Wew \leq 100$, implying a link to the DR phenomenon: in a drag-reducing turbulent flow, the viscoelasticity would work as a damping force against the streamwise vortex.
For even higher $\Wew=1000$ and 2000 cases, we had confirmed by longer time simulation runs for over $t^{\ast}=2000$ that the amplitudes of the periodic motion did not converge to a constant value. 
This indicates that the viscoelasticity may give rise to additional instabilities at high-enough Weissenberg numbers.
Similar Weissenberg-number effects were also observed in the Taylor--Couette flow~\citep{baumert97,crumeyrolle02}. 
The onset of this unsteadiness, or the pulsation, should be relevant to the route to EIT.
In addition to the pulsatile motion, the $K^\prime$ profile for $\Wew=2000$ showed a kink at every peak during the periodic state, which was not observed in the lower $\Wew$ cases.
This kink must be a result of the elasticity-induced instability; another investigation, however, will be required with a DNS at a much higher Weissenberg number.

\begin{figure}[t]
\begin{center}
	\includegraphics[width=0.75\hsize]{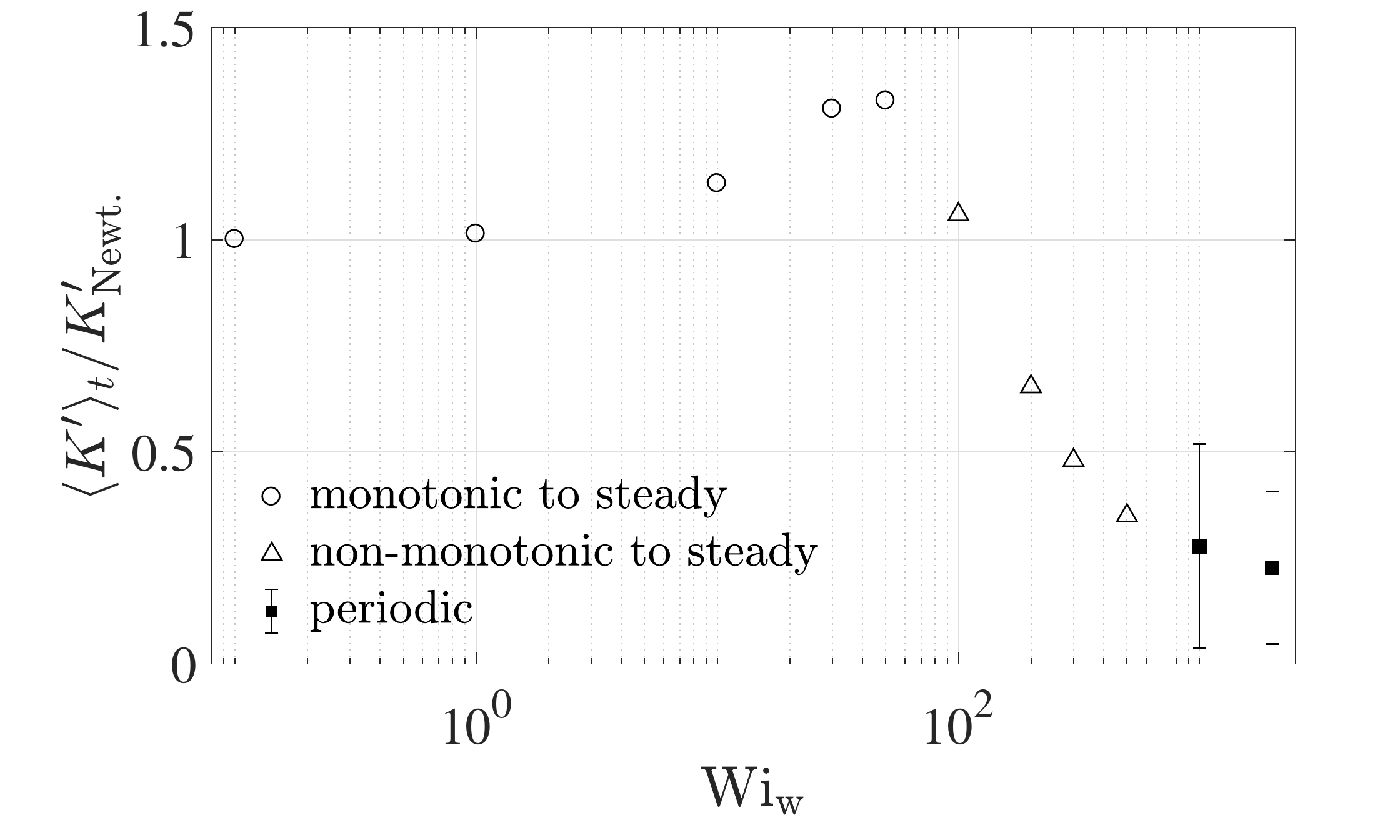}
	\caption{Dependency of $\langle K^\prime \rangle_t$ in the fully developed regime on the Weissenberg number. The values in the viscoelastic fluids were normalized by the Newtonian value. An error bar indicates the maximum and minimum values during a periodic behavior of the pulsatile flow. }
	\label{fig:K_wi}
\end{center}
\end{figure}

Figure~\ref{fig:K_wi} illustrates the dependence of $K^{\prime}$ in the fully developed regime on $\Wew$, where the mean values $\langle K^\prime \rangle_t$ averaged after the initial growth phase are plotted as a function of the Weissenberg number. 
The different symbols in the figure represent different flow regimes, and, here, we differentiated three regimes depending on the flow state reached after a long enough time had elapsed: the `monotonic to steady' case, where the flow reaches a steady state monotonically after an exponential growth; the `nonmonotonic to steady' case, where the flow exhibits an overshoot or where periodic behaviors appear but eventually decay before a steady state is reached; and the `periodic' case. 
For $\Wew \leq 50$, $\langle K^\prime \rangle_t$ increased with increasing $\Wew$. 
For larger $\Wew$ cases, where the $K^\prime$ evolutions showed the overshoot or the periodic behaviors, the opposite tendency was observed, i.e., the mean value $\langle K^\prime \rangle_t$ decreased as $\Wew$ further increased.

\begin{figure}[t]
\begin{center}
	\includegraphics[width=0.7\hsize]{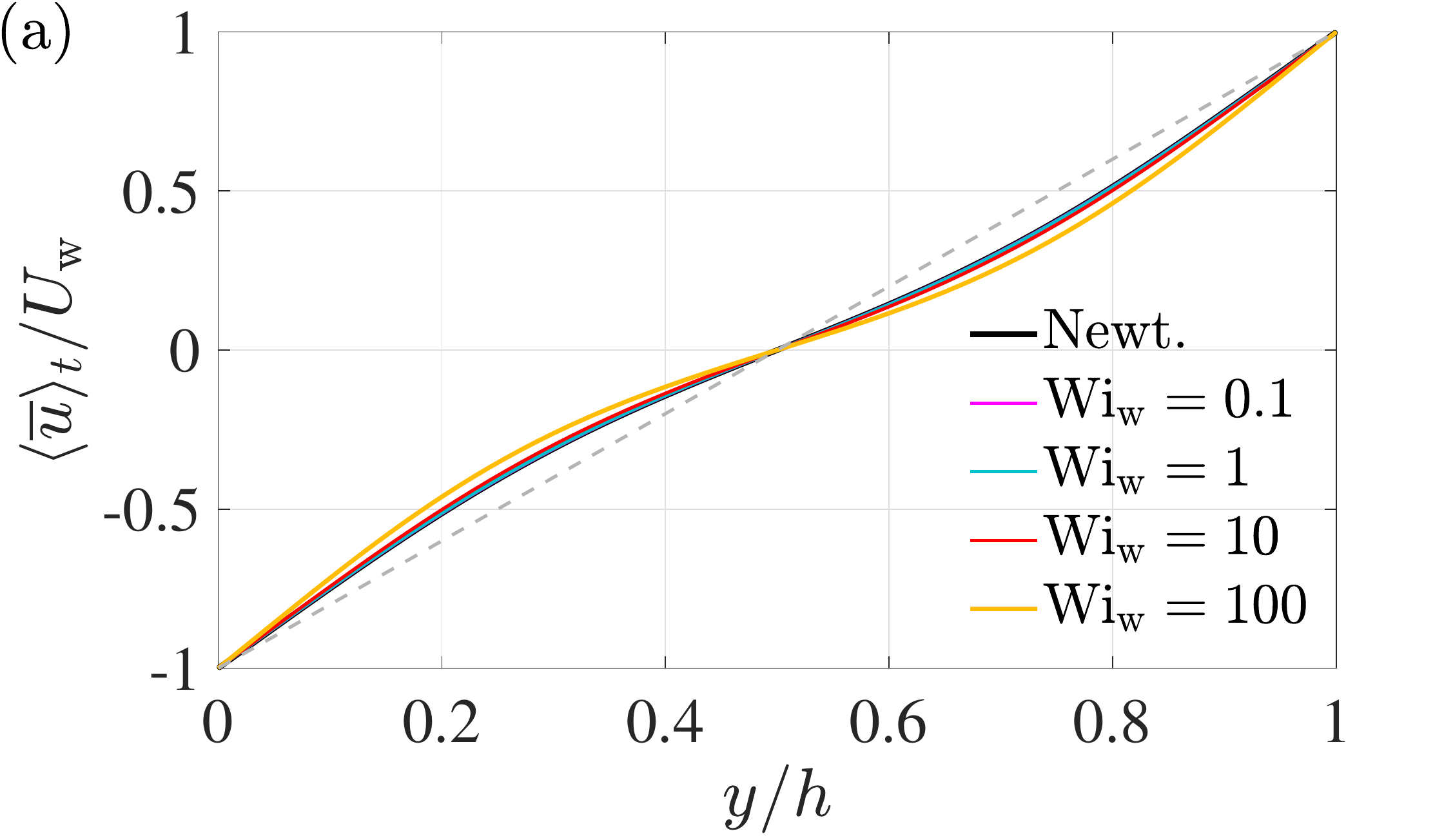}
	\includegraphics[width=0.7\hsize]{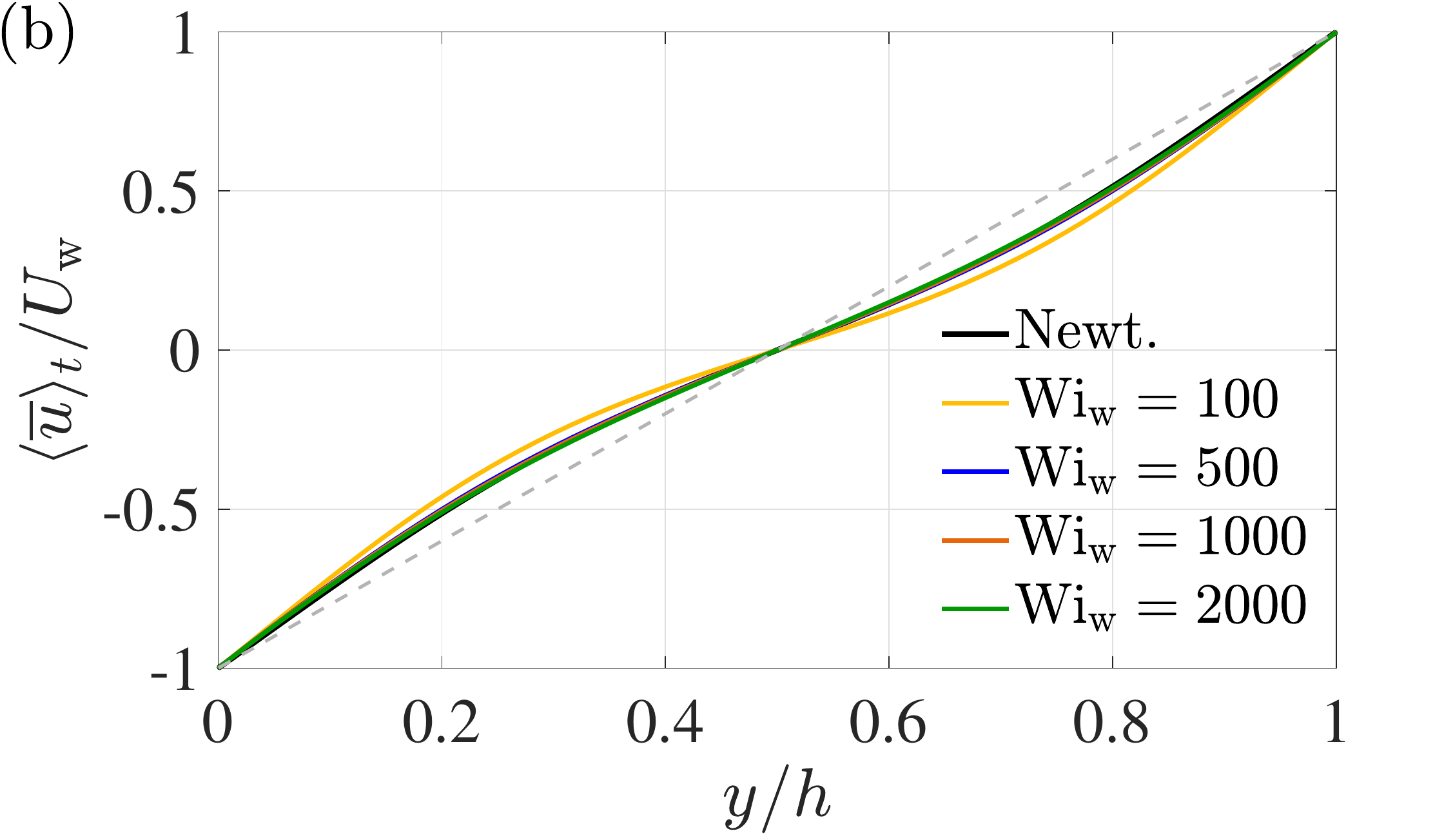}
	\caption{Profiles of the mean streamwise velocity averaged in the $x$- and $z$-directions and in time for different Weissenberg numbers with fixed $\Rew=25$ and $\Omega=10$. Panel~(a) includes the case in which the growth of $K^\prime$ exhibited a monotonic/nonmonotonic to a steady state, whereas panel~(b) includes the cases showing the growth from a nonmonotonic to a steady/periodic state.}
	\label{fig:mean_u}
\end{center}
\end{figure}

Figure \ref{fig:mean_u} shows the profiles of the mean streamwise velocity averaged in the $x$- and $z$-directions and in time after the flow reached the steady state or the amplitude of the periodic behavior converged to a constant value. In all cases, the mean streamwise velocity profiles somewhat deviated from the linear profile of the laminar base flow, and no significant difference was found between them despite the wide range of Weissenberg numbers investigated.

\begin{figure}[t]
\begin{center}
\includegraphics[width=0.7\hsize]{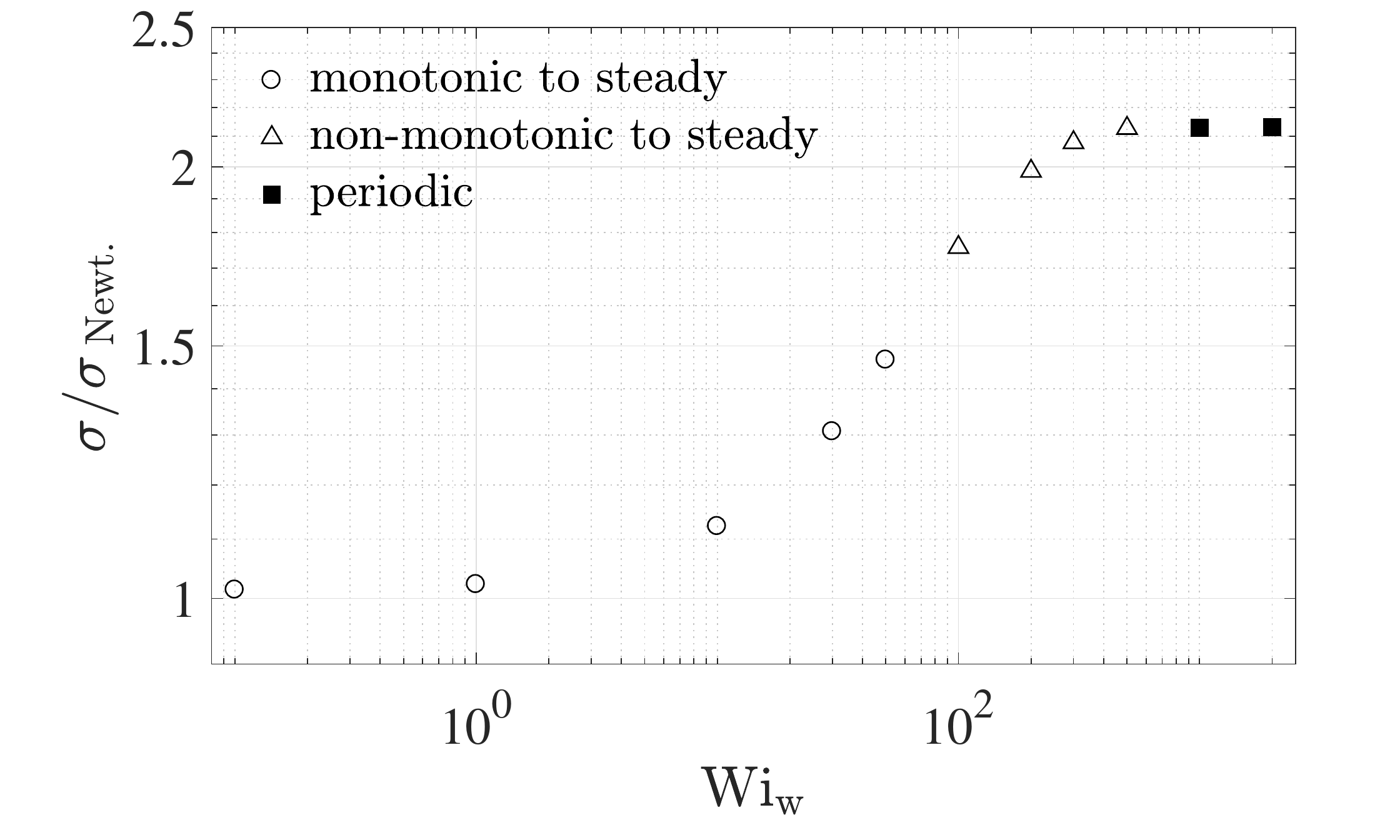}
\caption{Dependency of the growth rate $\sigma$ on the Weissenberg number. The values in the viscoelastic fluids were normalized by the Newtonian value. }
\label{fig:sigma}
\end{center}
\end{figure}

As shown in Fig.~\ref{fig:K_t}, the initial disturbances given to the flow grew exponentially, and the growth rates $\sigma=\mathrm{d} \ln K^\prime/\mathrm{d} t$ are evaluated in Fig.~\ref{fig:sigma} as the ratio to the Newtonian-case value. 
The same symbols as in Fig.~\ref{fig:K_wi} were used here.
It is shown that, for the low Weissenberg number range ($\Wew \leq 10$), the $\Wew$ dependency of the growth rate was not so strong, whereas in the middle range ($10 < \Wew < 500$) the growth rate significantly increased with increasing $\Wew$ and the `non-monotonic' state appeared. 
At even higher $\Wew$ cases, the flow exhibited the periodic state and the $\Wew$ dependency of the growth rate was no longer observed. It is particularly noteworthy that the $\Wew$ effect started to appear from $\Wew=1$, where the relaxation time of the additive was comparable to the viscous time scale of the flow. 
It is interesting to note that such viscoelasticity effect to increase the disturbance growth rate has also been reported by \cite{biancofiore17}, although the mechanisms of disturbance growth studied in this work were different from those of the present flow system; \cite{biancofiore17} studied the transient algebraic disturbance growth in a linearly stable plane Couette flow without system rotation, whereas in the present study the RPCF, where disturbance grows exponentially owing to a linear instability, was investigated. 

As shown above, at $\Wew=100$, the time evolution of $K^\prime$ had an overshoot before reaching the steady state, which is a sign of the periodic state at higher Weissenberg numbers. 
Instead of $\Wew$, the additive relaxation time $\lambda$ can be non-dimensionalized as
\begin{equation}
 {\rm De} \equiv \frac{\lambda}{\delta / U_{\rm w}} = \frac{\rm Wi_w}{\Rew},
\label{eq:De}
\end{equation}
using the laminar shear rate $\delta / U_{\rm w}$.
For the present case, the Deborah number is the unity at $\Wew=25$, above which an instability would be induced by the viscoelasticity rather than by the inertia.
This is consistent with the abovementioned sign of unsteadiness, or the periodic state, observed for $\Wew \geq 100$, which corresponds to ${\rm De} \geq 4$.
As discussed later, the onset of the steadily pulsatile flow is related to the turnover time scale of the roll cell and of the system rotation. 

Further details of the steady and pulsatile flow states at lower and higher Weissenberg numbers are described in Sections~\ref{steady_flow} and \ref{unsteady_flow}, respectively. 

\subsection{Steady flow fields}
\label{steady_flow}

\begin{figure}
\begin{center}
	\includegraphics[width=0.7\hsize]{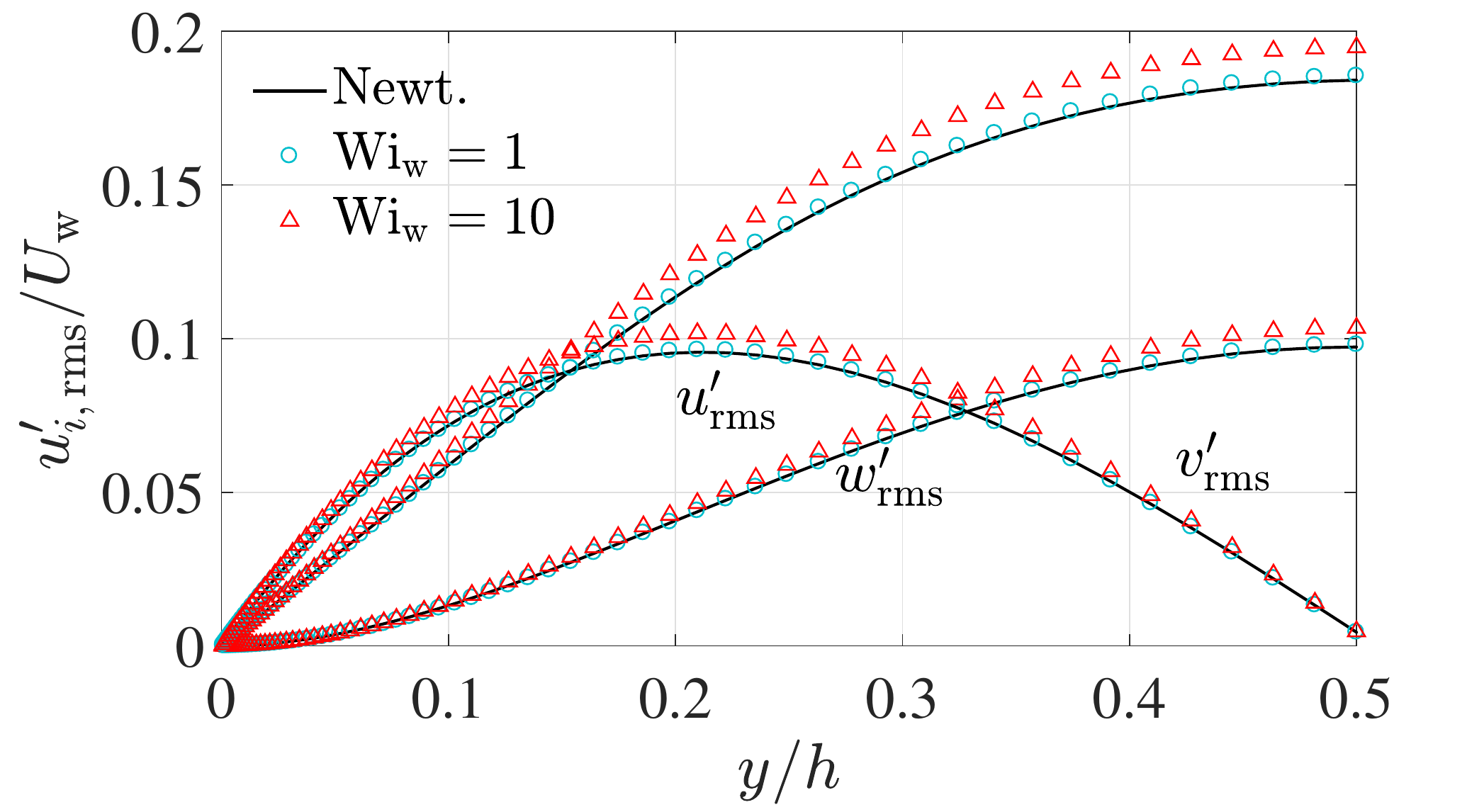}
	\caption{Wall-normal profiles of the velocity variation intensity normalized by $U_{\rm w}$.}
	\label{fig:rms}
\end{center}
\end{figure}

As described in the previous section, the flow reached a steady state after an exponential disturbance growth, and the $\langle K^\prime \rangle_t$ level at the steady state was enhanced as $\Wew$ increased up to $\Wew=50$. Figure~\ref{fig:rms} shows the profiles of the root-mean-square of the velocity deviation for the Newtonian, $\Wew = 1$, and $\Wew = 10$ cases as the representative cases of the steady flow states. 
Here, $u^\prime_\mathrm{rms}$, $v^\prime_\mathrm{rms}$, and $w^\prime_\mathrm{rms}$ were evaluated by averaging in the $x$- and $z$-directions at the last time step of each case.
It can be seen in the figure that no components were significantly affected by $\Wew$, although they slightly increased with increasing $\Wew$, indicating that the vortical motion of the streamwise-independent straight roll cells observed in the Newtonian case was slightly enhanced by the addition of viscoelasticity for this relatively low $\Wew$ range. 

\subsection{Unsteady flow fields}
\label{unsteady_flow}

\begin{figure*}[t]
\begin{center}
	\includegraphics[width=0.99\hsize]{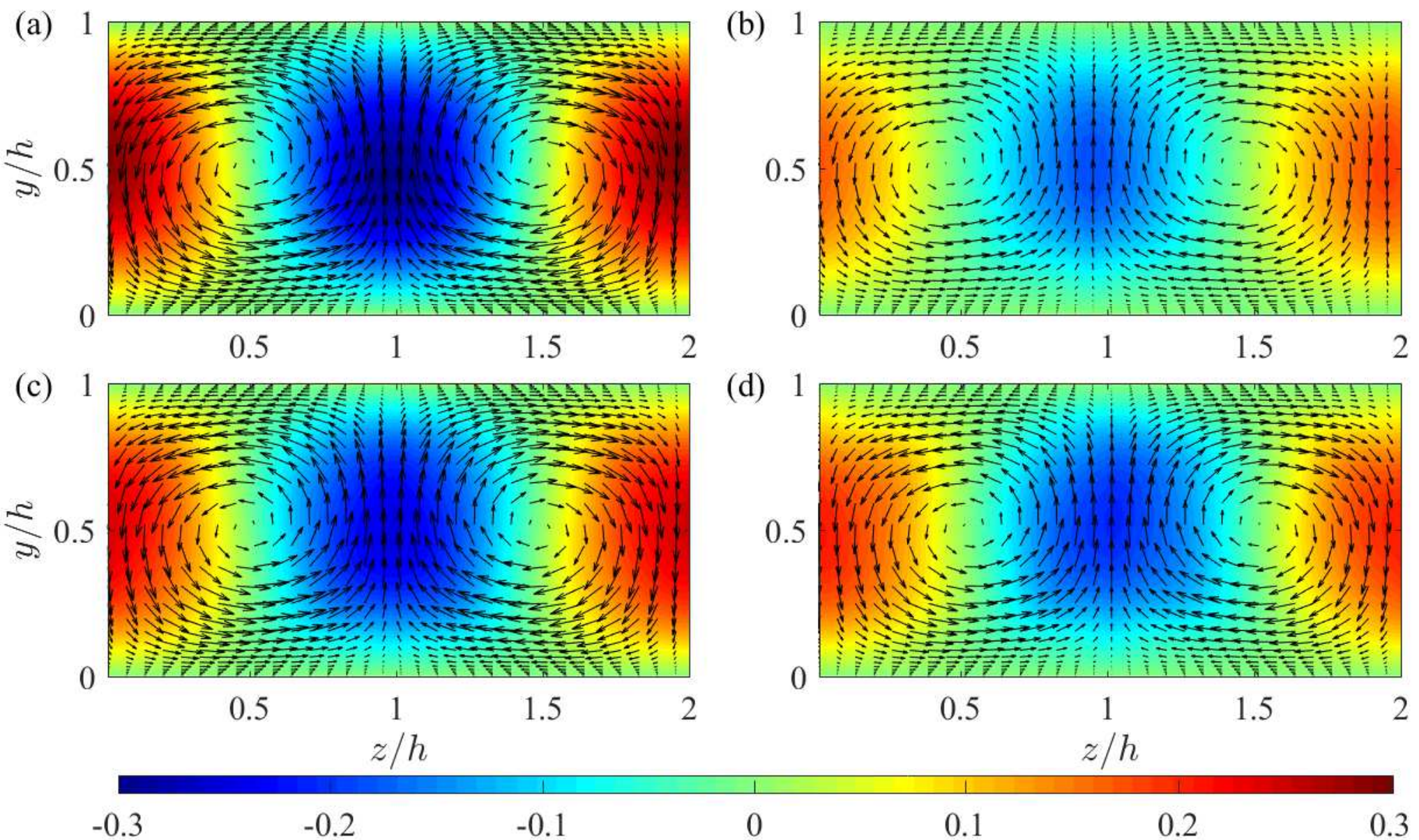}\\
	\caption{Comparison of the roll-cell structure of the RPCF at $\Rew=25$ and $\Omega=10$ for (a) the Newtonian fluid and (b--d) the viscoelastic fluid with three different values of $\Wew$: (b) $\Wew=500$, (c) $\Wew=1000$, and (d) $\Wew=2000$. 
The colors indicate the streamwise velocity $u^{\prime}/U_{\rm w}$, and the black arrows show the pattern of the cross-flow vector $(v,w)$. The arrow with $2h$ length corresponds to the velocity magnitude of $U_{\rm w}$. 
Snapshots of (c, d) show each field at an instance giving the maximum $K^\prime$ during its pulsatile motion.}
	\label{fig:rolls}
\end{center}
\end{figure*}

\begin{figure}[t]
\begin{center}
	\includegraphics[width=0.7\hsize]{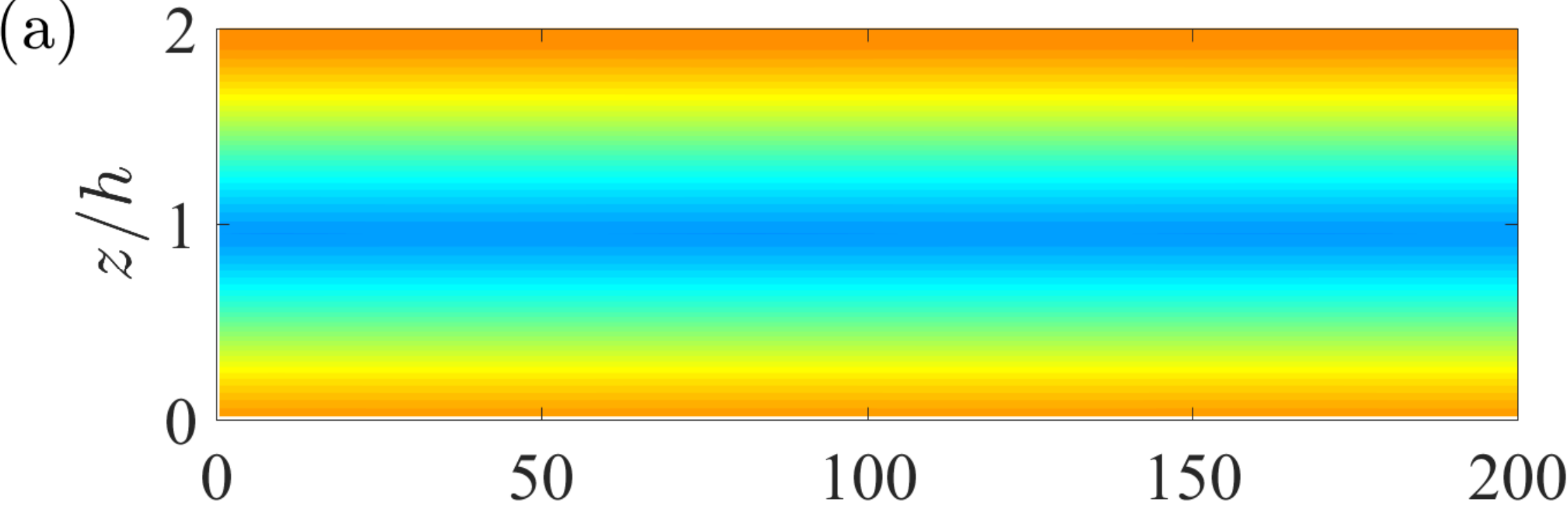}\\
	\includegraphics[width=0.7\hsize]{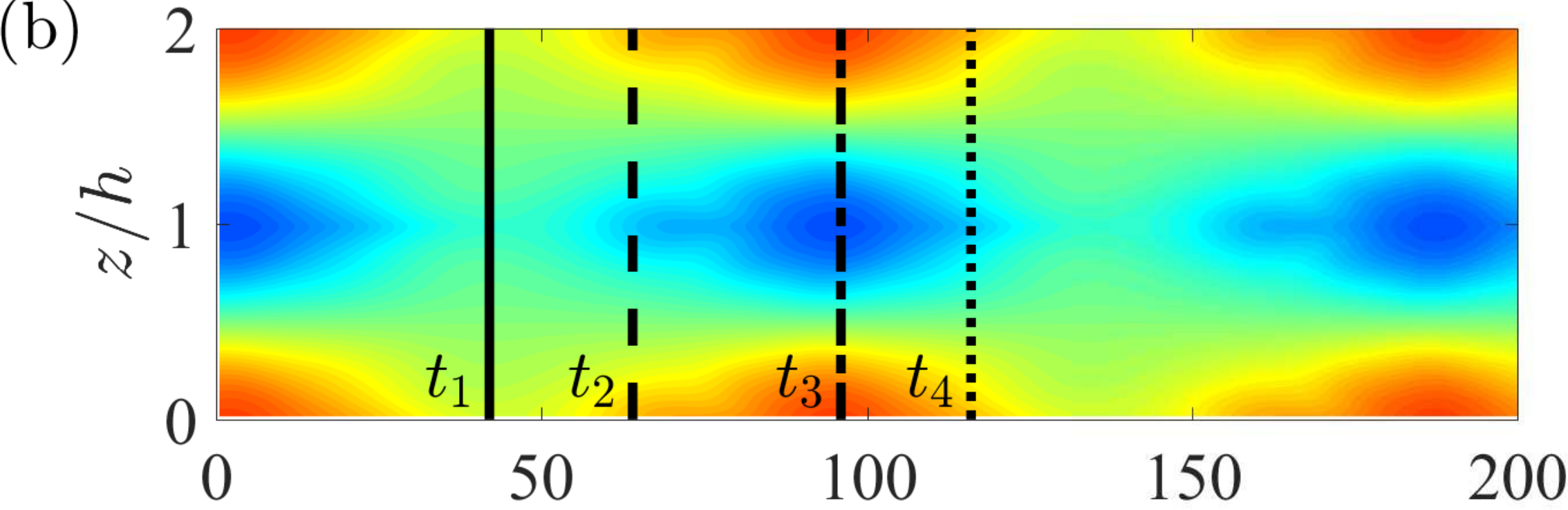}\\
	\includegraphics[width=0.7\hsize]{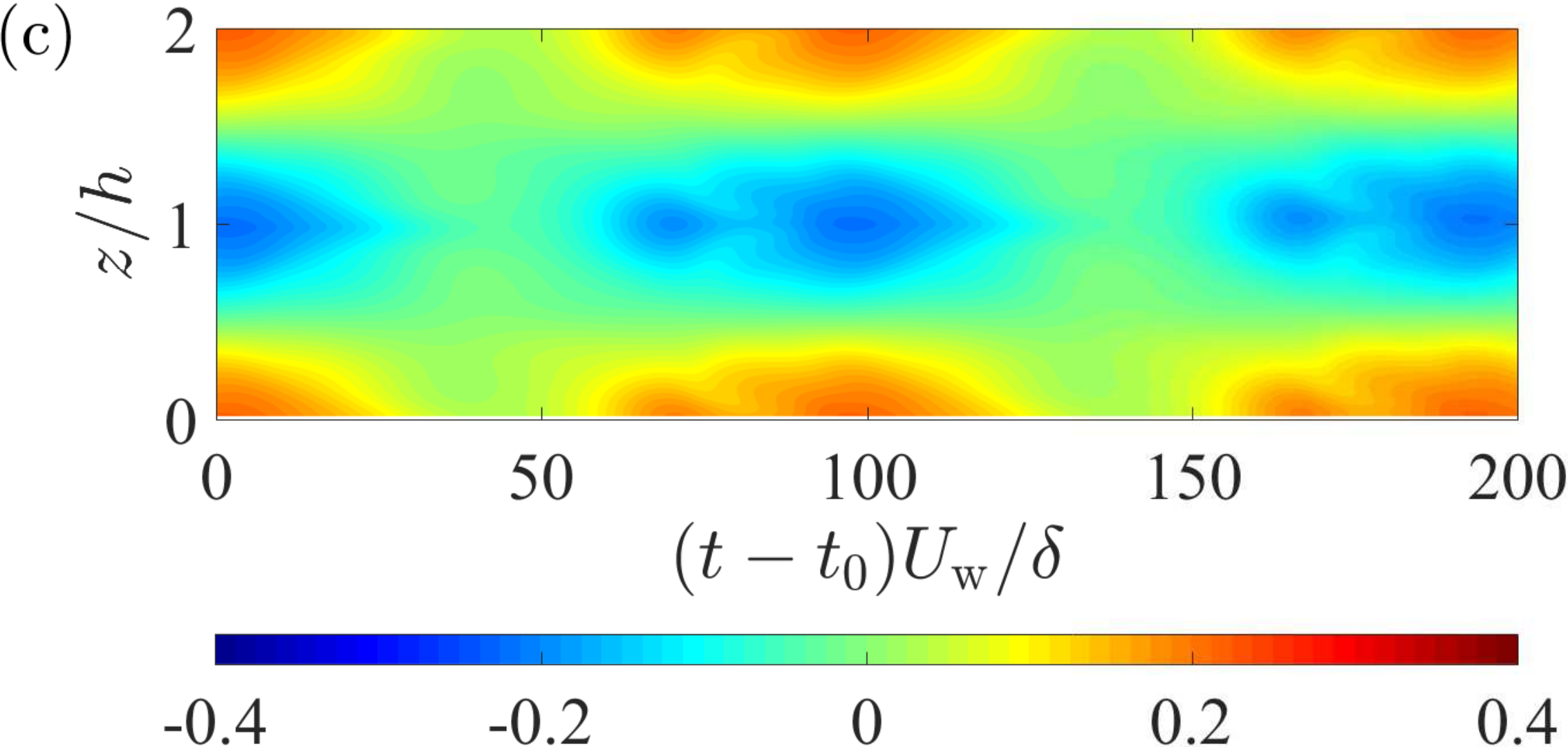}\\
	\caption{Space--time diagrams of the streamwise velocity $u^{\prime}/U_{\rm w}$ on the channel centerline $y/h=0.5$ for the viscoelastic fluid cases: (a) $\Wew=500$, (b) $\Wew=1000$, and (c) $\Wew=2000$. 
The black lines in panel (b) represent the instances depicted for later analysis. }
	\label{fig:stdiag}
\end{center}
\end{figure}

\begin{figure}[t]
\begin{center}
	\includegraphics[width=0.7\hsize]{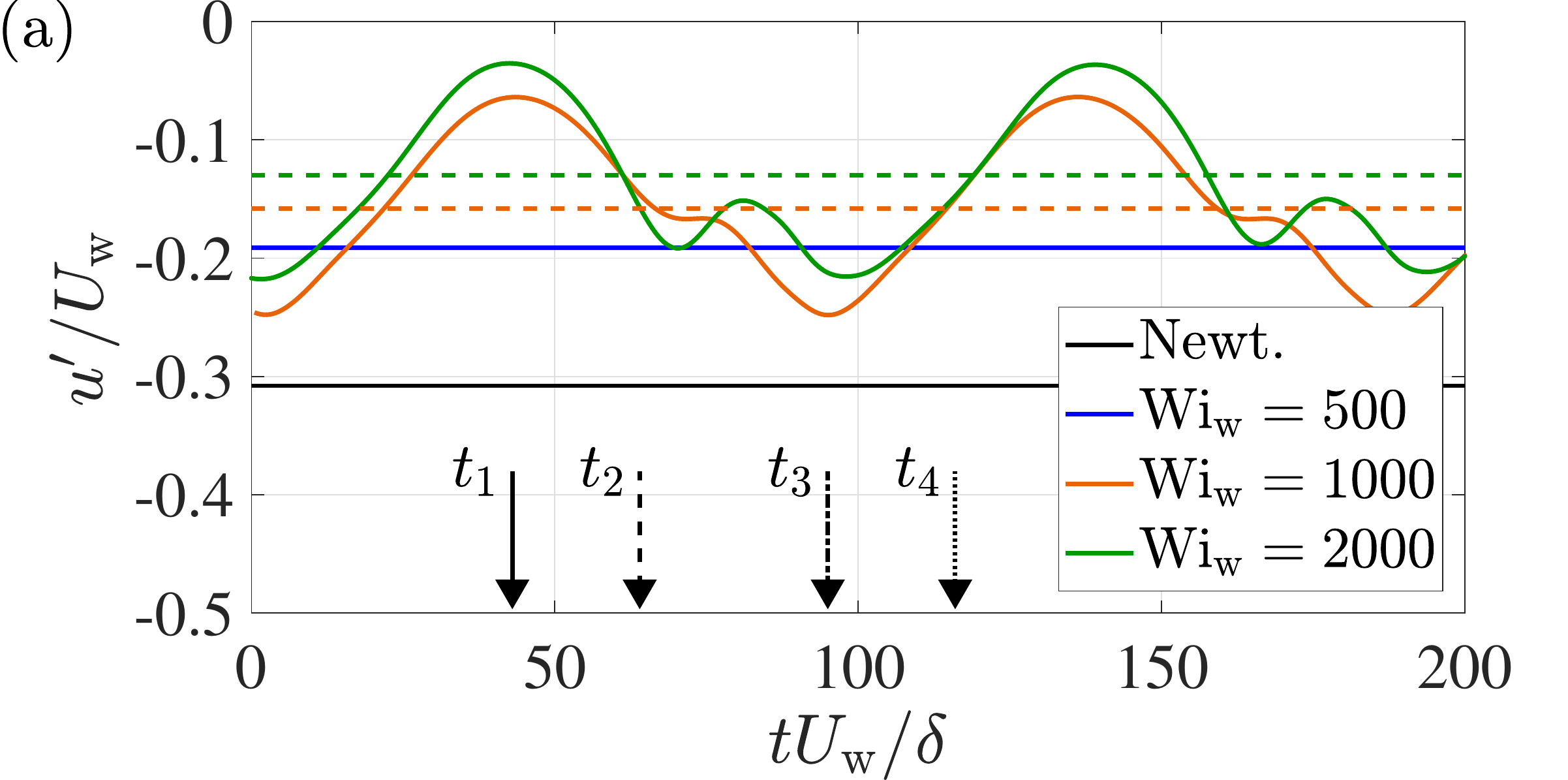}\\
	\includegraphics[width=0.7\hsize]{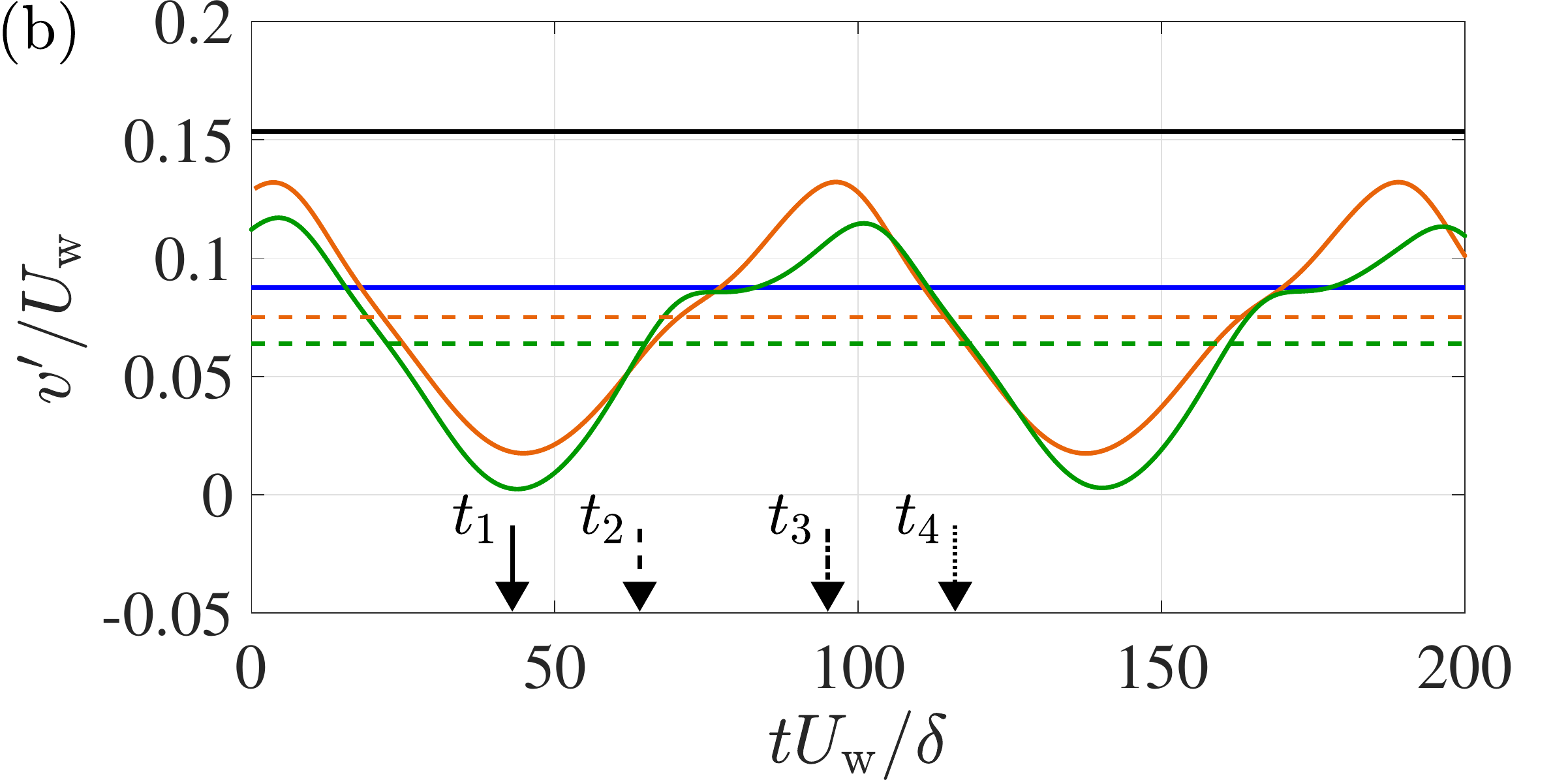}
	\caption{Time sequence of the variations in (a) the streamwise velocity $u^{\prime}/U_\mathrm{w}$ and (b) the wall-normal velocity $v^{\prime}/U_\mathrm{w}$ at the channel center ($y/h=0.5$ and $z/h=1.0$), shown in solid lines. Each of the four solid lines indicates a Newtonian fluid and different values of $\Wew$. Dashed lines indicate the time-averaged values $\langle u'/U_\mathrm{w} \rangle_t$ and $\langle v'/U_\mathrm{w} \rangle_t$ at $(y/h=0.5,z/h=1.0)$ for the $\Wew=1000$ and $2000$ cases. The four time instances of $t_1$--$t_4$ correspond to those in Fig.~\ref{fig:stdiag}. }
	\label{fig:opm}
\end{center}
\end{figure}

Figure~\ref{fig:rolls} shows the cross-sectional views of the flow structures observed in the laminar RPCFs of the Newtonian and viscoelastic fluids for each $\Wew$. 
The contour indicates the magnitude of the streamwise velocity component $u^\prime/U_\mathrm{w}$, and the arrows represent the pattern of the cross-flow vectors $(v^\prime,w^\prime)$. 
Note that the mean components of $v$ and $w$ are zero across the channel in the present flow system and, therefore, $(v,w)=(v^\prime,w^\prime)$. 
As shown in the figure, a large-scale roll-cell structure with the size of the channel gap clearly appeared for all cases, and roll cells induced the periodic spanwise variation in the streamwise velocity of approximately $15\%$ of the wall speed $U_\mathrm{w}$. 
It should also be noted that all these observed structures were two-dimensional, i.e., homogeneous in the streamwise direction. 
Although the spatial structures that appeared in these four cases were quite similar to each other, a significant difference was observed in their temporal behaviors. 
Figure~\ref{fig:stdiag} shows the space--time diagrams ($z$--$t$ diagram) of $u^{\prime}/U_\mathrm{w}$ on the centerline of the channel at an arbitrary streamwise position. 
Only the viscoelastic fluid cases of the fully developed states after the transient regime ($t > t_0 \gg 0$) are shown. 
From the diagram, one may find a pair of positive and negative regions that alternate in the spanwise direction. 
These regions corresponded to the streaks induced by the streamwise roll cells. 
For $\Wew=500$ as well as the Newtonian fluid, the 2D roll cells were steady, as given in Fig.~\ref{fig:stdiag}(a), where the position and magnitude of the streaks did not change in time. 
Although Fig.~\ref{fig:K_t} reveals a transiently unsteady $K^\prime$ for $\Wew=500$, its flow eventually became steady as given in  Figs.~\ref{fig:stdiag}(a) and \ref{fig:opm}, with $t_0 \gg 500\delta/U_{\rm w}$ (discussed again later).
At the higher $\Wew$ of Figs.~\ref{fig:stdiag}(b) and (c), the periodic variations in the streamwise velocity were clearly confirmed. 
It is also shown here that, despite the significant variation in the magnitude, the velocities did not change their sign: for example, at $z/h=1$ in panel (b), $u^\prime$ was always negative, whereas the magnitude significantly changed in time. 
Therefore, the flow structure only repeated being enhanced and damped periodically without altering the sign or position.
This periodic feature of roll cells affected by elasticity is called `pulsatile motion,' hereinafter.

More details on such `pulsatile motion' of the structure are presented in Fig.~\ref{fig:opm}, which compares the time series of the magnitude of the streamwise and wall-normal velocities at $y/h=0.5$ and $z/h=1.0$ of Fig.~\ref{fig:stdiag} over a period of $200 U_\mathrm{w}/\delta$ from an arbitrary time $t_0$ after the flow reached a statistically steady state. 
As shown in the figure, $u^\prime$ and $v^\prime$ were constant in time in the Newtonian case and those at $\Wew=500$ also exhibited the same pattern, although they were oscillating with quite a small amplitude. 
The periodic variation of $K^\prime$ at the early stage of the simulation of the $\Wew=500$ case shown in Fig.~\ref{fig:K_t} gradually decayed in time, and the amplitude shown in Fig.~\ref{fig:opm} was negligibly small. However, in the higher-$\Wew$ cases, significant periodic variations could be seen. The amplitude of the pulsatile motion was approximately $10\%$ and $5\%$ of the wall speed $U_\mathrm{w}$ for $u^\prime$ and $v^\prime$, respectively, at both $\Wew=1000$ and 2000. Although the temporal changes in $u^\prime$ and $v^\prime$ were monotonic with respect to their damping phases (around $t_4$), the enhancement phase (around $t_2$) exhibited a non-monotonic aspect. 
Such a kink in the periodic variation was also seen in the time variation of $K^\prime$, as shown in Fig.~\ref{fig:K_t}(b). 

As shown in Fig.~\ref{fig:opm}, the period of the pulsatile motion $t_P$ was significantly long compared to the inverse of the shear rate, and the pulsatile periods for these cases were evaluated as $t_P=92.8\delta/U_\mathrm{w}$ and $92.4\delta/U_\mathrm{w}$ at $\Wew=1000$ and 2000, respectively. 
As the value of $\lambda / (\delta / U_\mathrm{w} )= {\rm De}$ for $\Wew=1000$ and $2000$ was 40 and 80, the periods of the pulsatile motion were, at least for these two $\Wew$ cases, on the same order as the relaxation time. 
It is interesting to note that, although the relaxation time was doubled between $\Wew=1000$ and 2000, a significant difference was not observed between the pulsatile-motion periods of these cases. 
It is also noteworthy here that these pulsation period were on the same order as the time scale of the hibernating turbulence observed by \cite{xi10}, who also observed that the time scale of the hibernation was less dependent on the Weissenberg number. 
The ratio of the relaxation time to the turnover time of the cell rotation could provide a critical value of the Weissenberg number, above which pulsation is observed. 
The cell turnover time $t_{\rm cell}$ may be quantified by the time-averaged rotation speed and radius of the vortex. 
Because the roll cell should be normalized with an outer-scale unit, both the cell rotation speed and the radius are respectively related to the wall speed and the gap width.
However, the viscoelasticity may reduce the speed more with increasing $\Wew$. 
According to the time-averaged values, shown as dashed lines in Fig.~\ref{fig:opm}, the magnitude of $\langle v' \rangle_t$, which represents the cell rotation speed, decreased as $\Wew$ increased. 
Employing $\langle v^\prime \rangle_t$ and $\delta$ as the characteristic values of cell rotation, we estimated $t_{\rm cell}= 88\delta/U_\mathrm{w}$ and $97\delta/U_\mathrm{w}$ for $\Wew=1000$ and 2000, respectively. 
The pulsation period could possibly be related to the turnover time of the cell flow, by which the pulsation period was scaled well rather than by the relaxation time. 

\begin{figure}[t]
\begin{center}
	\includegraphics[width=0.7\hsize]{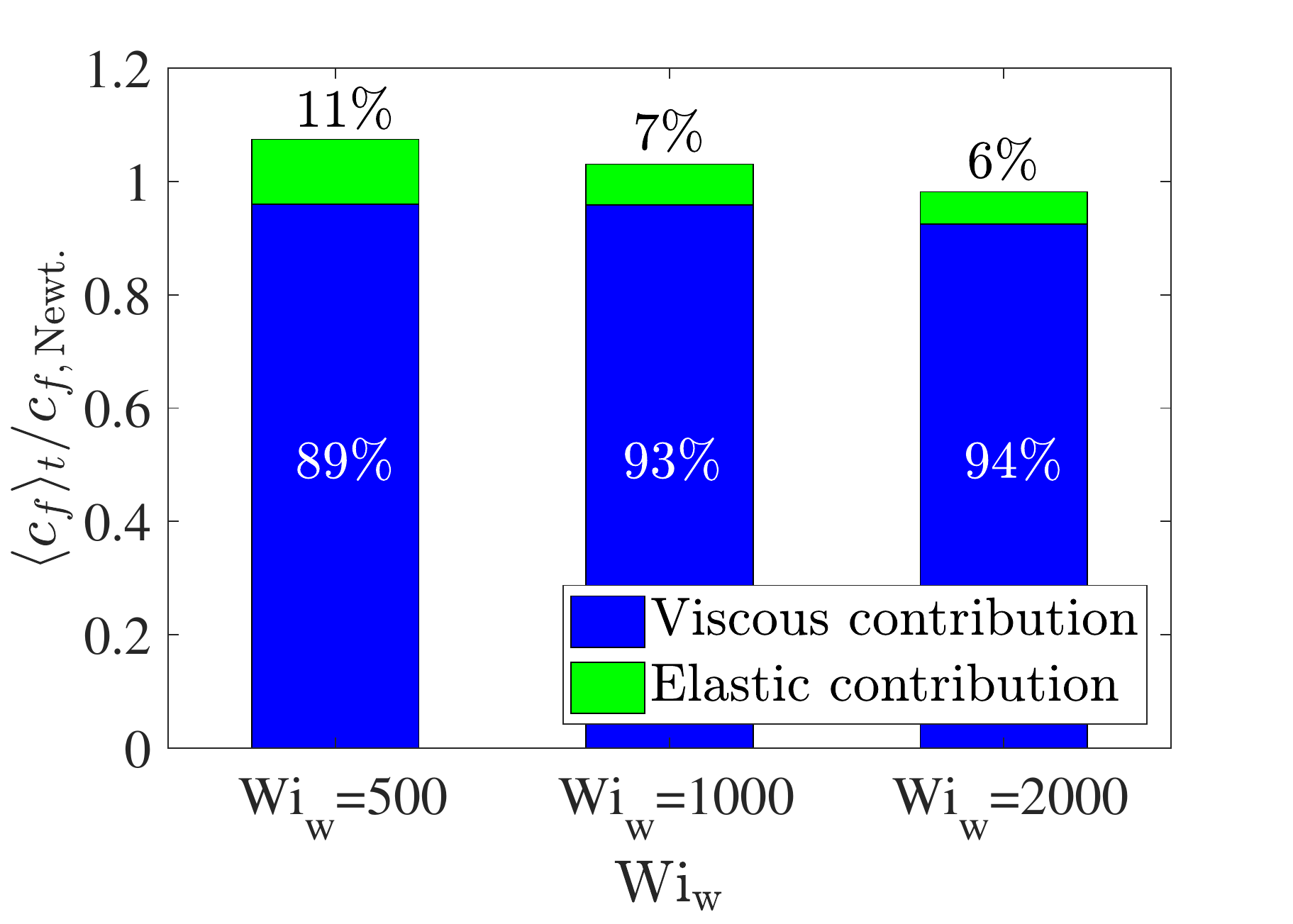}
	\caption{Proportion of the viscous and elastic contribution to skin friction coefficient compared to the value of the Newtonian fluid ($C_{f{\rm , Newt.}}$) in each of the Weissenberg-number cases.}
	\label{fig:cf_bar}
\end{center}
\end{figure}

Given the viscoelasticity, the spatial velocity variations caused by the roll cell were attenuated, as mentioned above. 
Figure~\ref{fig:cf_bar} shows the variation in the skin-friction coefficient as a function of $\Wew$. 
The values of $\langle c_f \rangle_t$ are shown as the ratio to the Newtonian case value, and it is shown that the viscous contribution decreased compared to the Newtonian case for all $\Wew$ cases; however, together with the elastic contribution, the overall skin friction increased at $\Wew=500$ and 1000, but decreased at $\Wew=2000$ compared to the Newtonian case.   
The decrease in the viscous contribution indicates that the momentum transport by the flow structure was weakened by the viscoelasticity, which was consistent with the tendency observed in the time series of Fig.~\ref{fig:opm}. 

\begin{figure}[t]
\begin{center}
	\includegraphics[width=0.7\hsize]{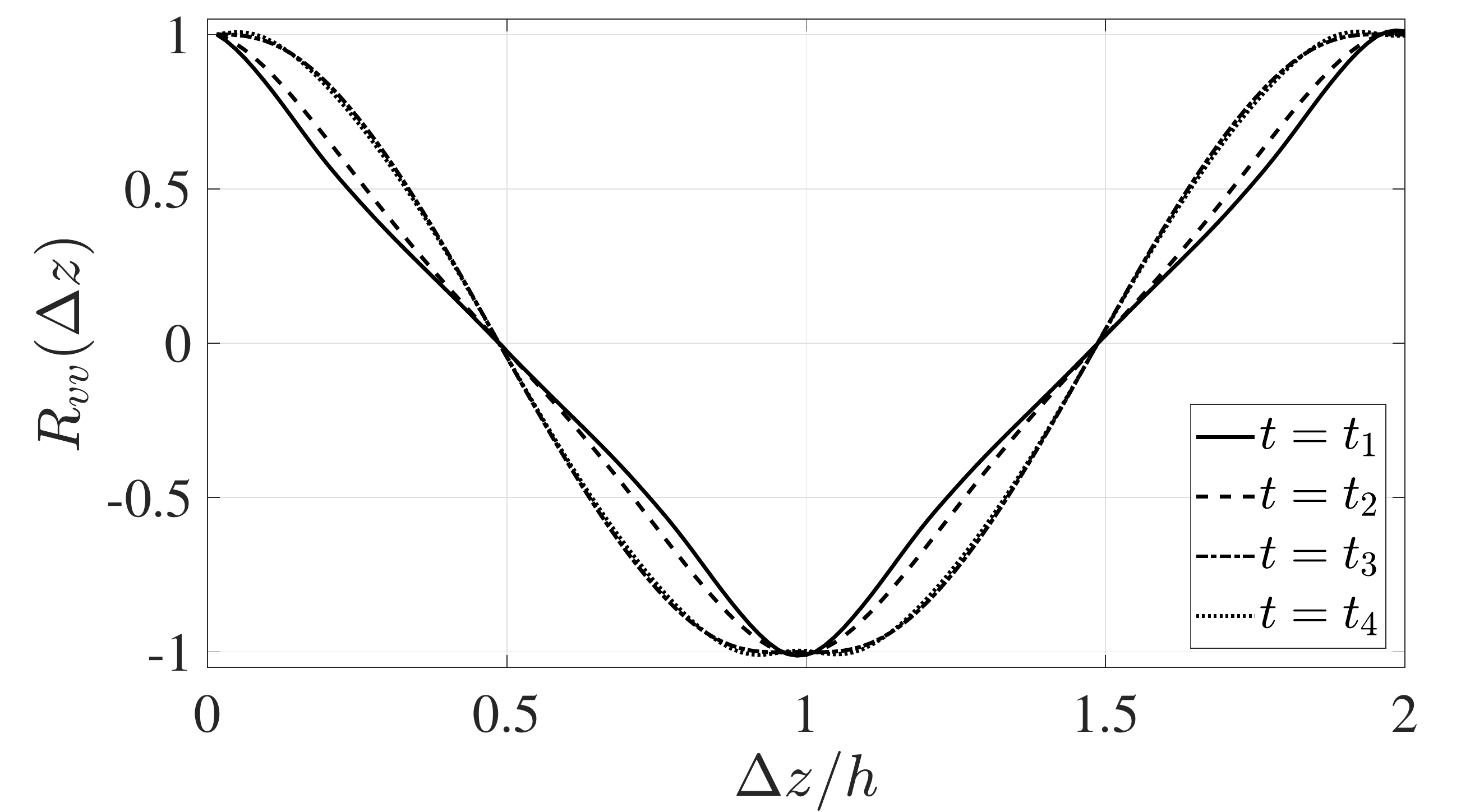}
	\caption{Spanwise spatial two-point correlation function $R_{vv} (\Delta z)$ of the wall-normal velocity on the channel centerline for $\Wew=1000$. The profiles were plotted for four different instances, i.e., $t=t_1$, $t_2$, $t_3$, and $t_4$, as shown in Fig.~\ref{fig:stdiag}(b) by the corresponding black lines.}
	\label{fig:rvv}
\end{center}
\end{figure}

Now, we select four representative instances from one period of the pulsatile motion for $\Wew=1000$, which are defined in Fig.~\ref{fig:stdiag}(b) with black vertical lines. 
The time instants of $t=t_1$ and $t_3$ are the moments when the spanwise variation in $u^{\prime}/U_\mathrm{w}$ (on the channel centerline) is the least and most significant, respectively; $t_2$ and $t_4$ are those when the trace of the conformation tensor $c_{xx} + c_{yy} + c_{zz}$ becomes minimum and maximum, respectively. 
Figure~\ref{fig:rvv} shows the spatial two-point correlation function $R_{vv}(\Delta z)$ of the wall-normal velocity $v^\prime(x,h/2,z)$ on the channel centerline at these moments: 
\begin{equation}
R_{vv}(\Delta z) = \frac{\overline{v^\prime(x,h/2,z)v^\prime(x,h/2,z+\Delta z)}}{v^\prime_{\rm rms}(h/2) v^\prime_{\rm rms}(h/2)},
\end{equation}
where the overline denotes the spatial averaging, and the profiles at four different instances are shown by the same lines as in Fig.~\ref{fig:stdiag}(b). The profile of $R_{vv}$ remains essentially unchanged throughout the pulsatile motion. In particular, the curves obtained at $t=t_3$ and $t_4$ overlap each other. This indicates that during the pulsatile motion only the magnitude of the vortical motion was varied in time, whereas the `shape' of the roll-cell structure, such as the spanwise width of the roll cells, was not significantly changed. 

\begin{figure*}[t]
 \begin{tabular}{ccc}
  \hspace{-0.9em}
  \begin{minipage}{0.32\hsize}
   \includegraphics[width=1\hsize]{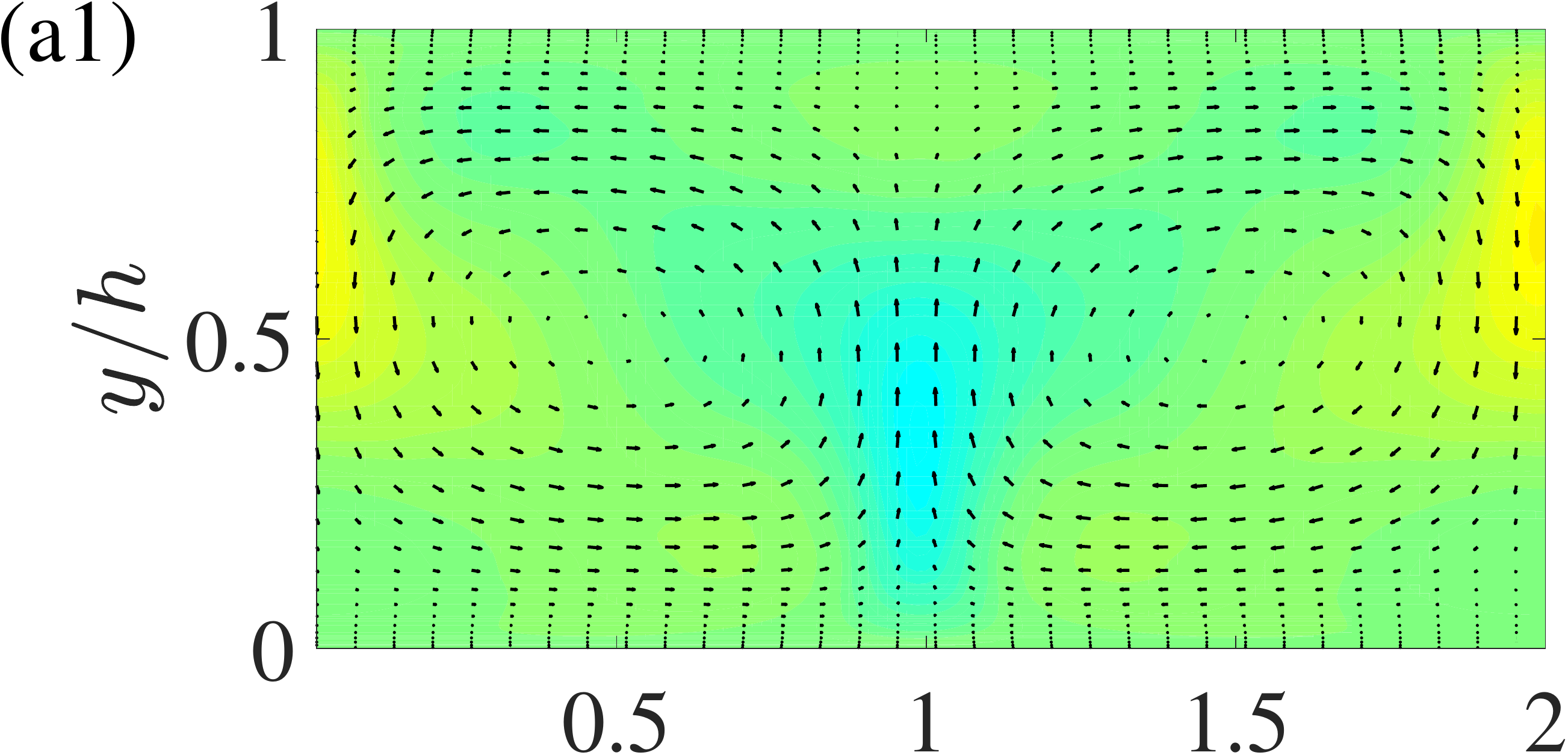}\\
   \includegraphics[width=1\hsize]{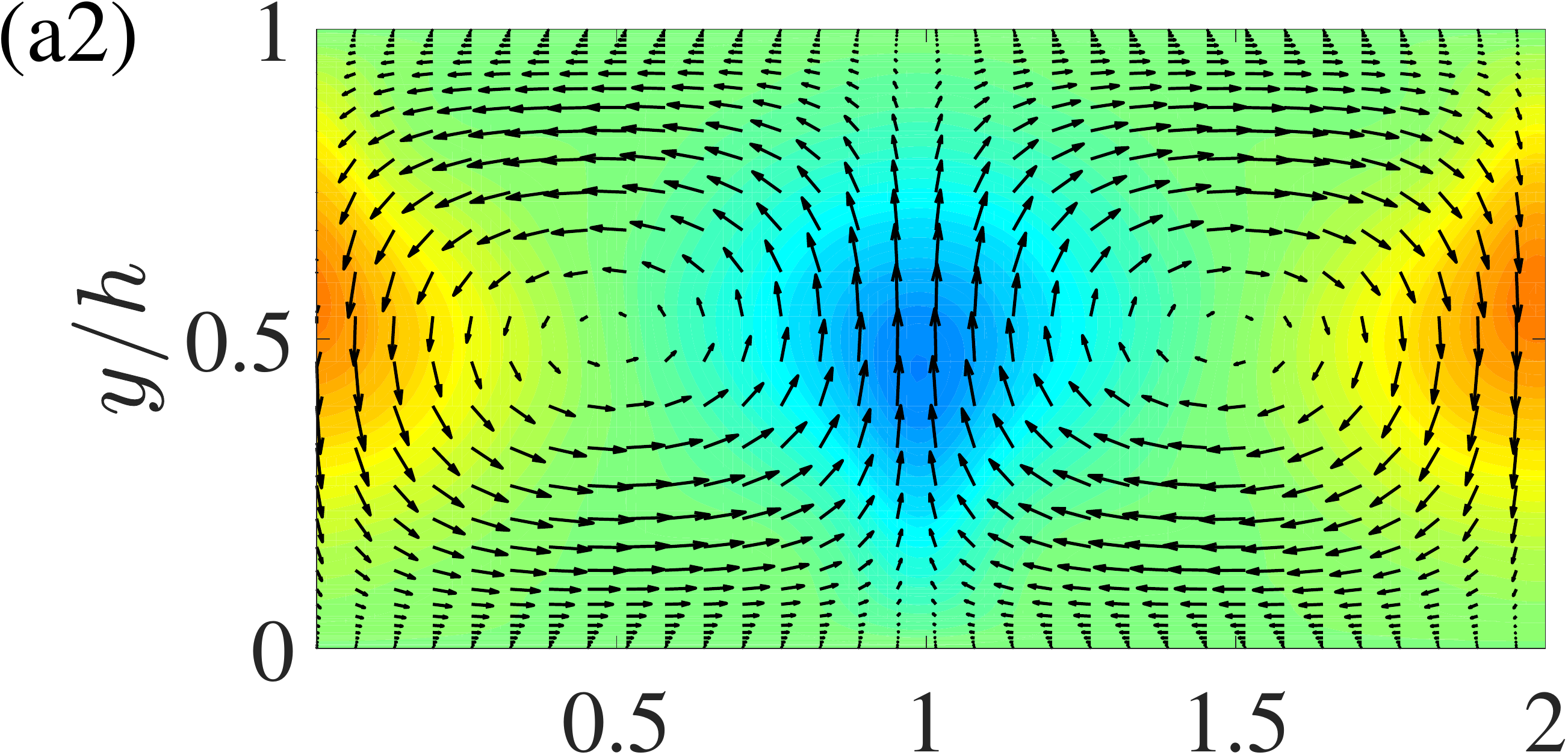}\\
   \includegraphics[width=1\hsize]{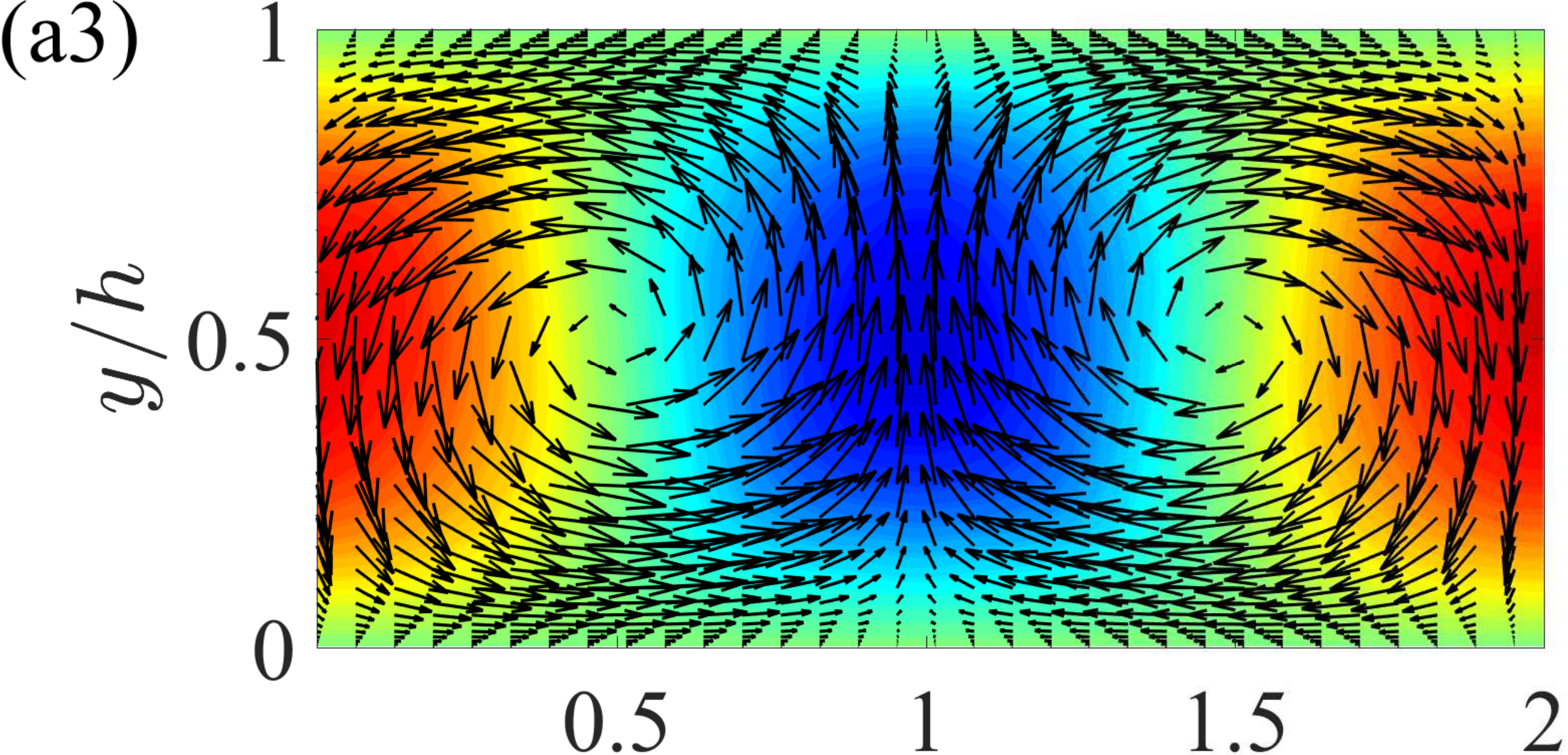}\\
   \includegraphics[width=1\hsize]{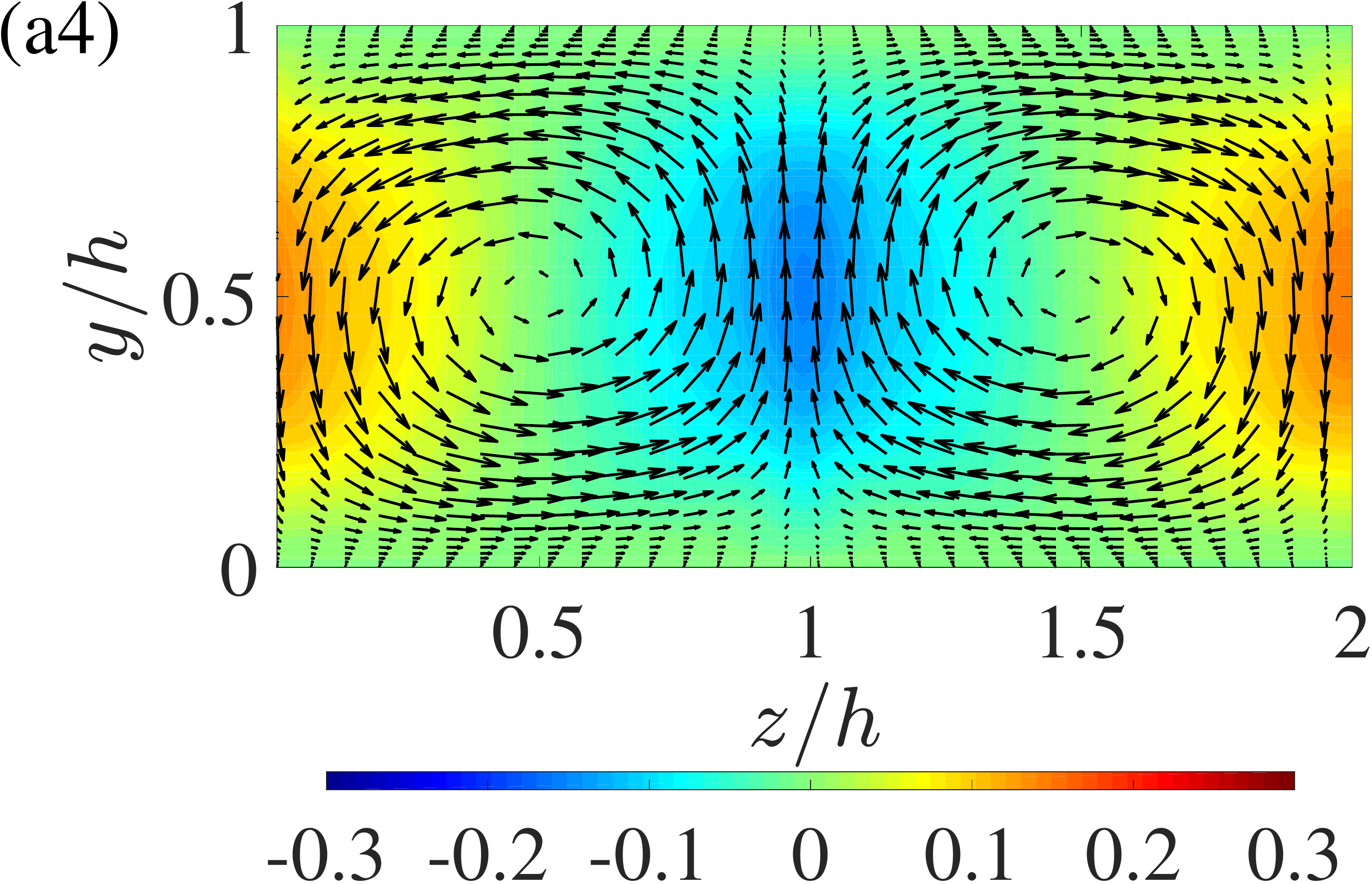}
  \end{minipage}
  \hspace{-0.35em}
  \begin{minipage}{0.32\hsize}
   \includegraphics[width=1\hsize]{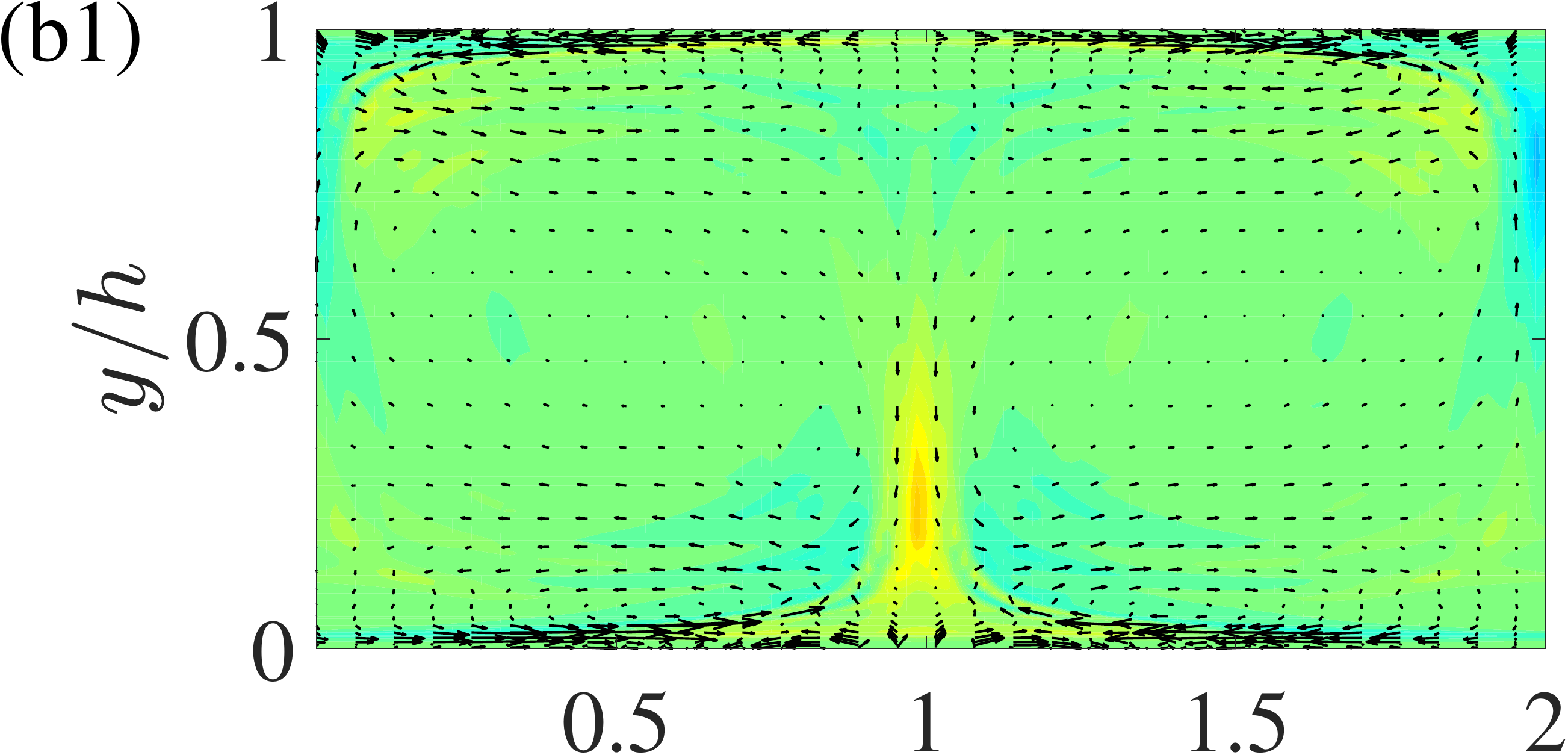}\\
   \includegraphics[width=1\hsize]{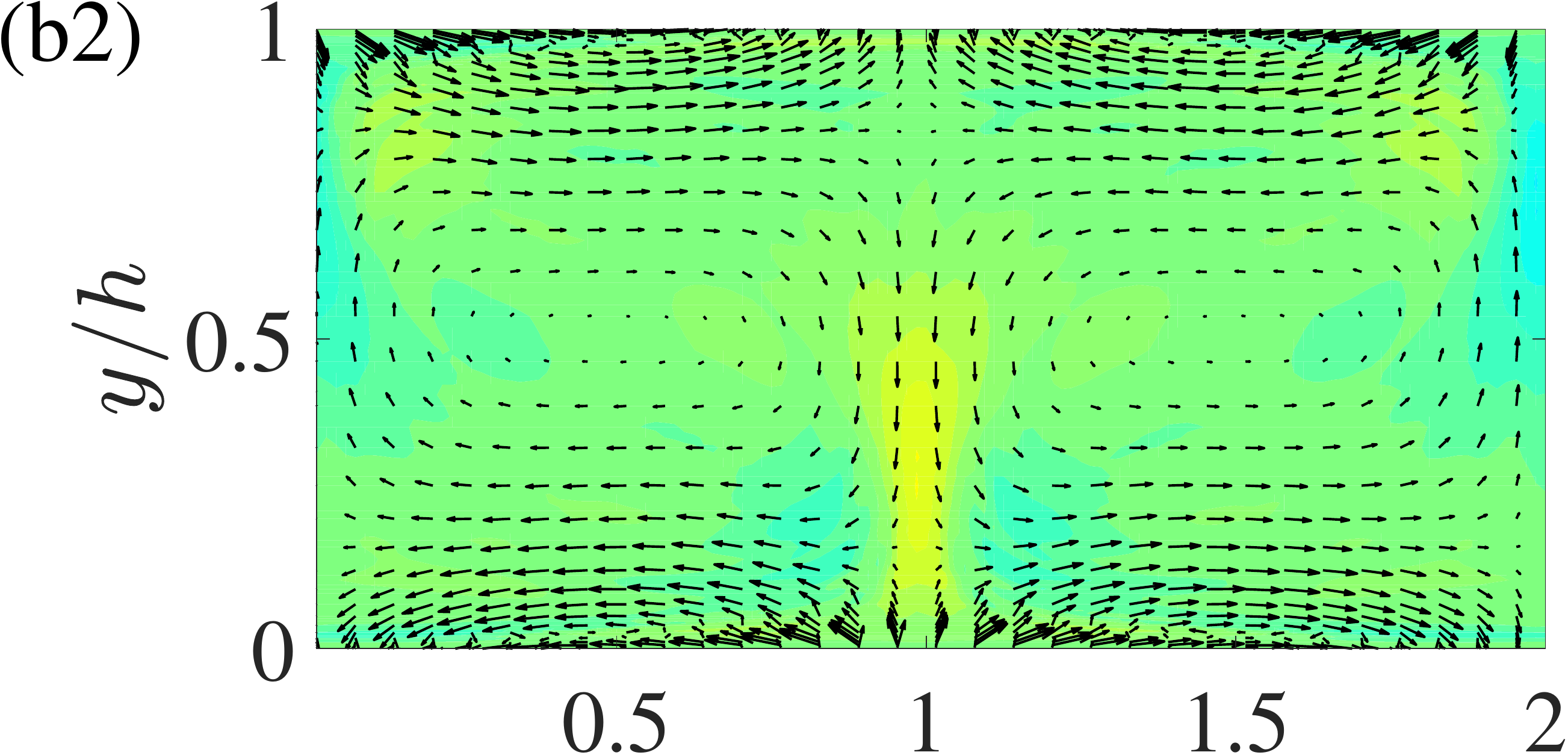}\\
   \includegraphics[width=1\hsize]{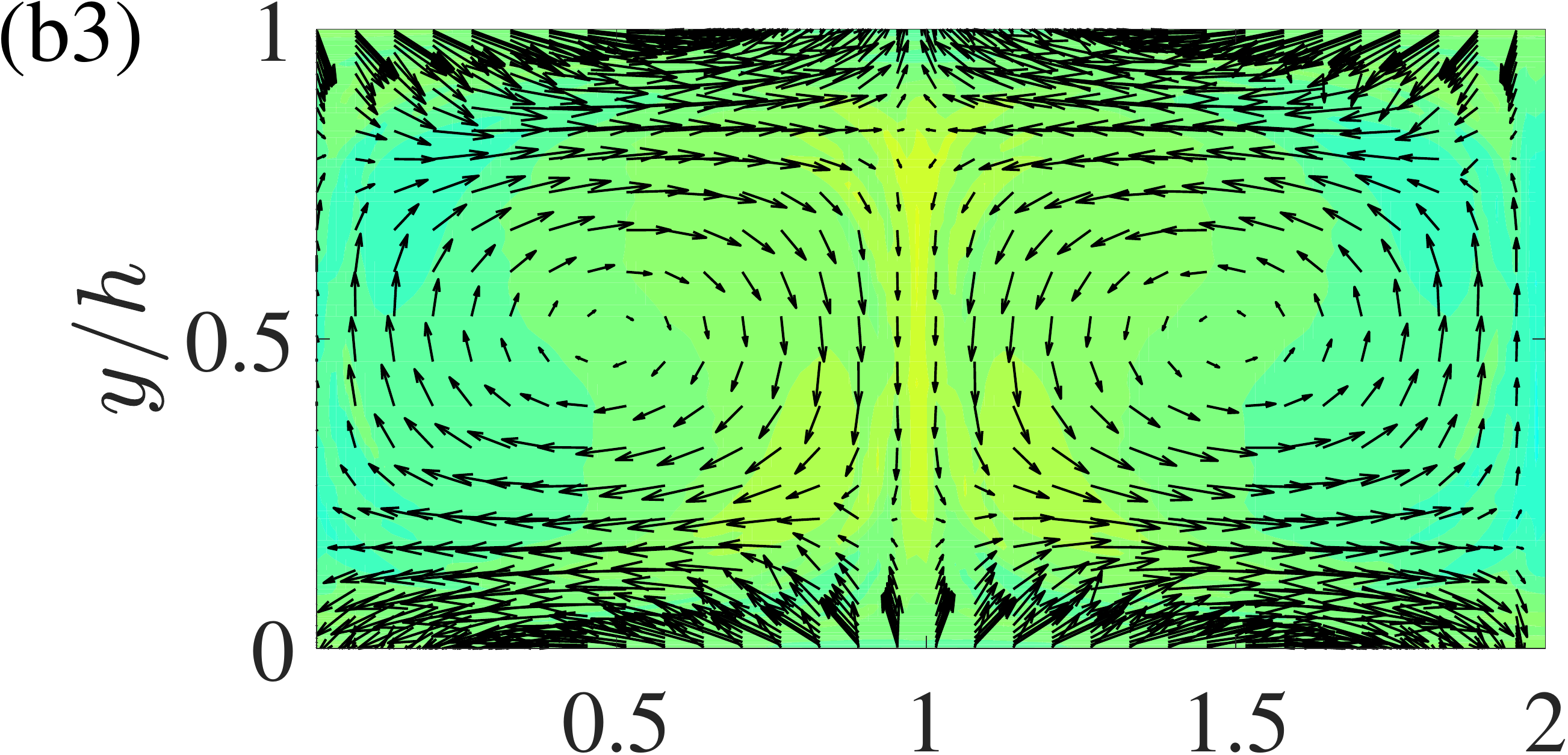}\\
   \includegraphics[width=1\hsize]{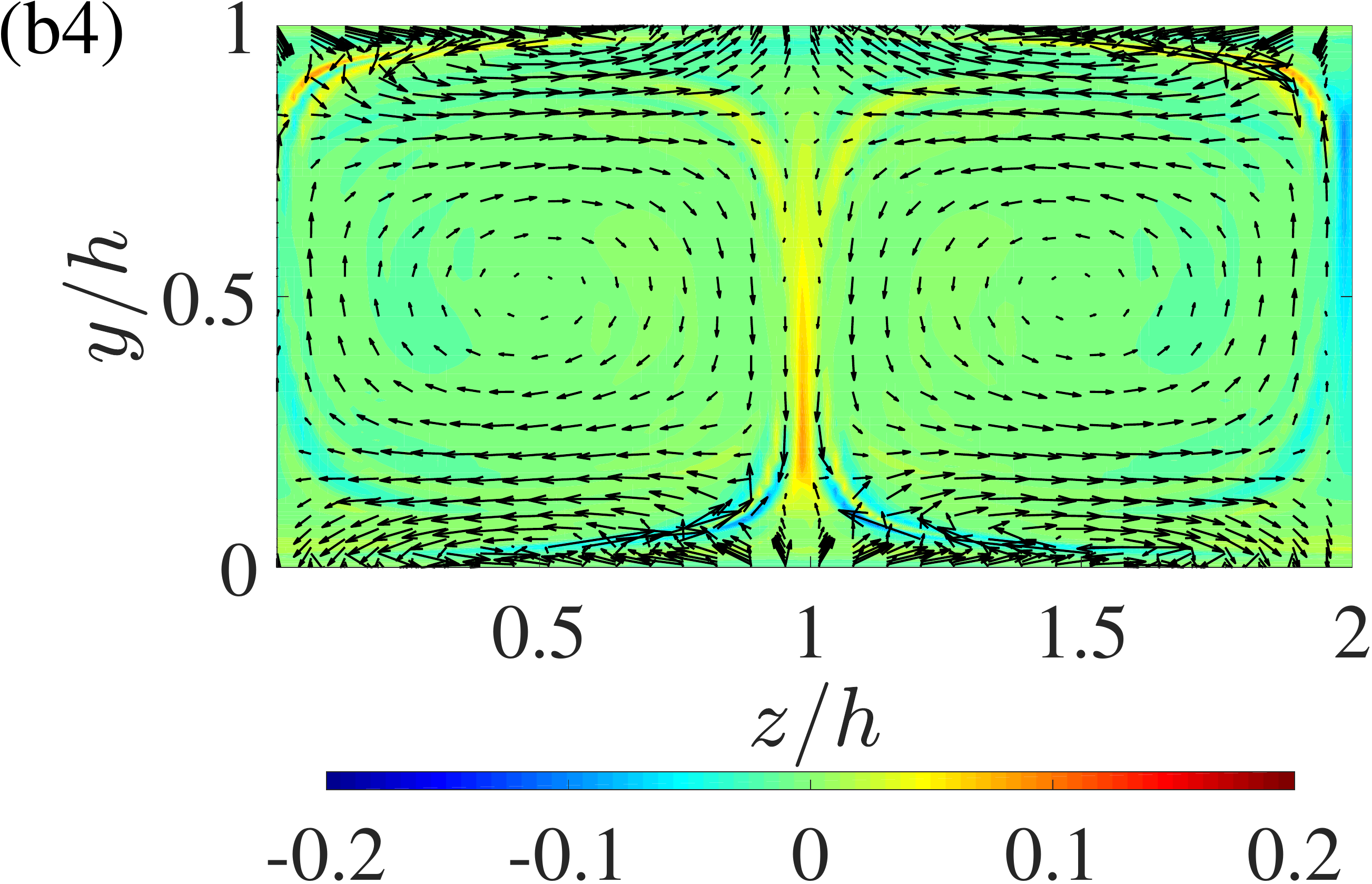}
  \end{minipage}
  \hspace{-0.35em}
  \begin{minipage}{0.32\hsize}
   \includegraphics[width=1\hsize]{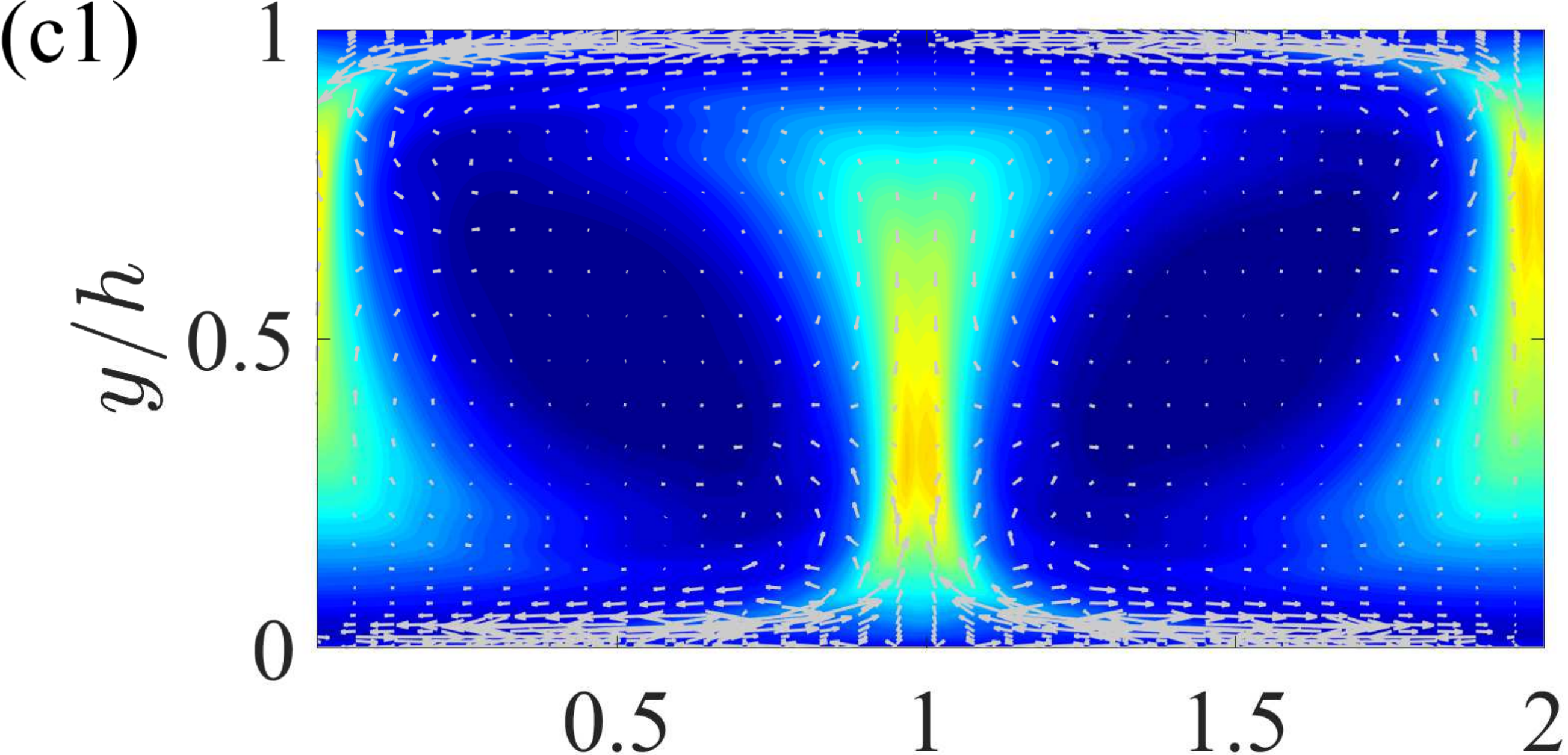}\\
   \includegraphics[width=1\hsize]{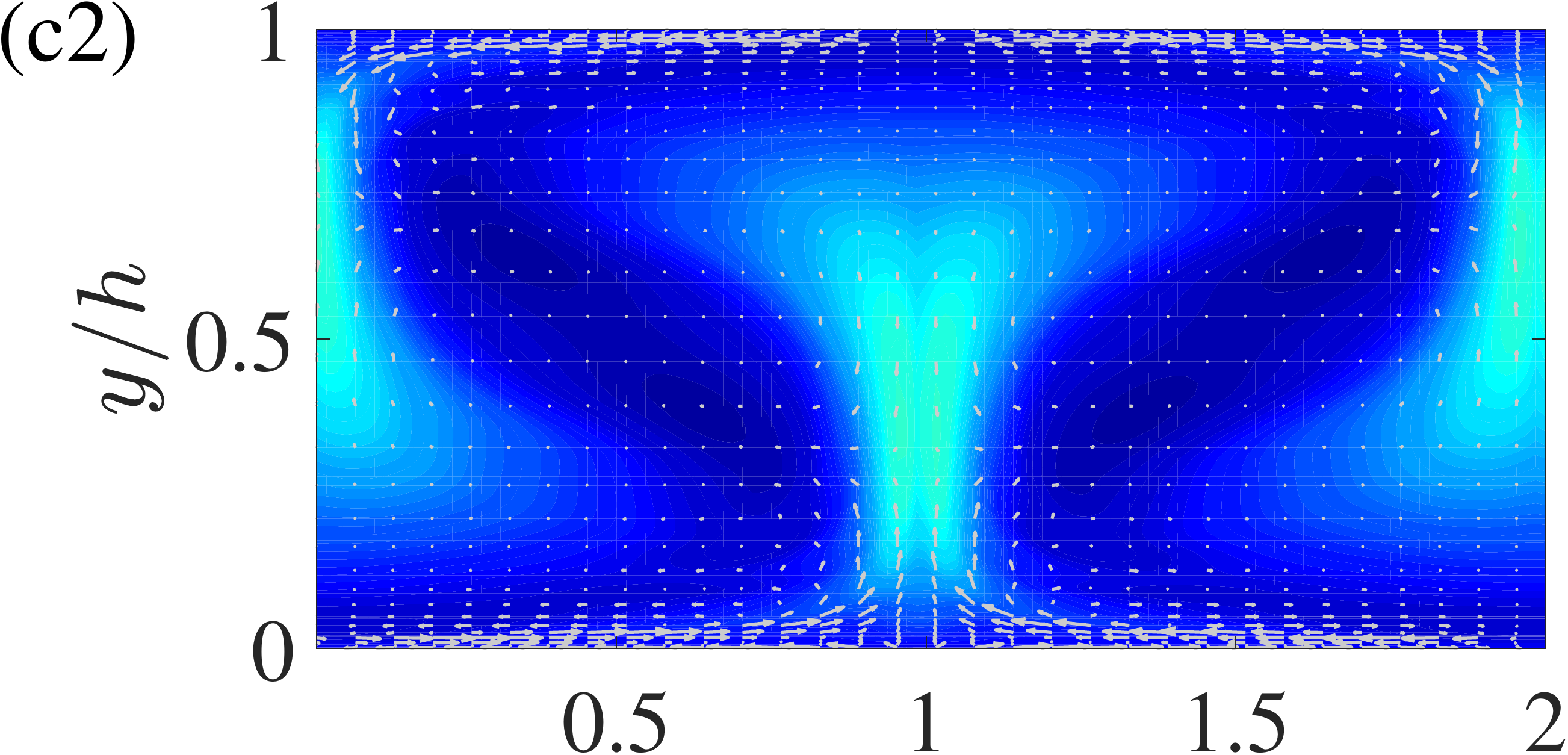}\\
   \includegraphics[width=1\hsize]{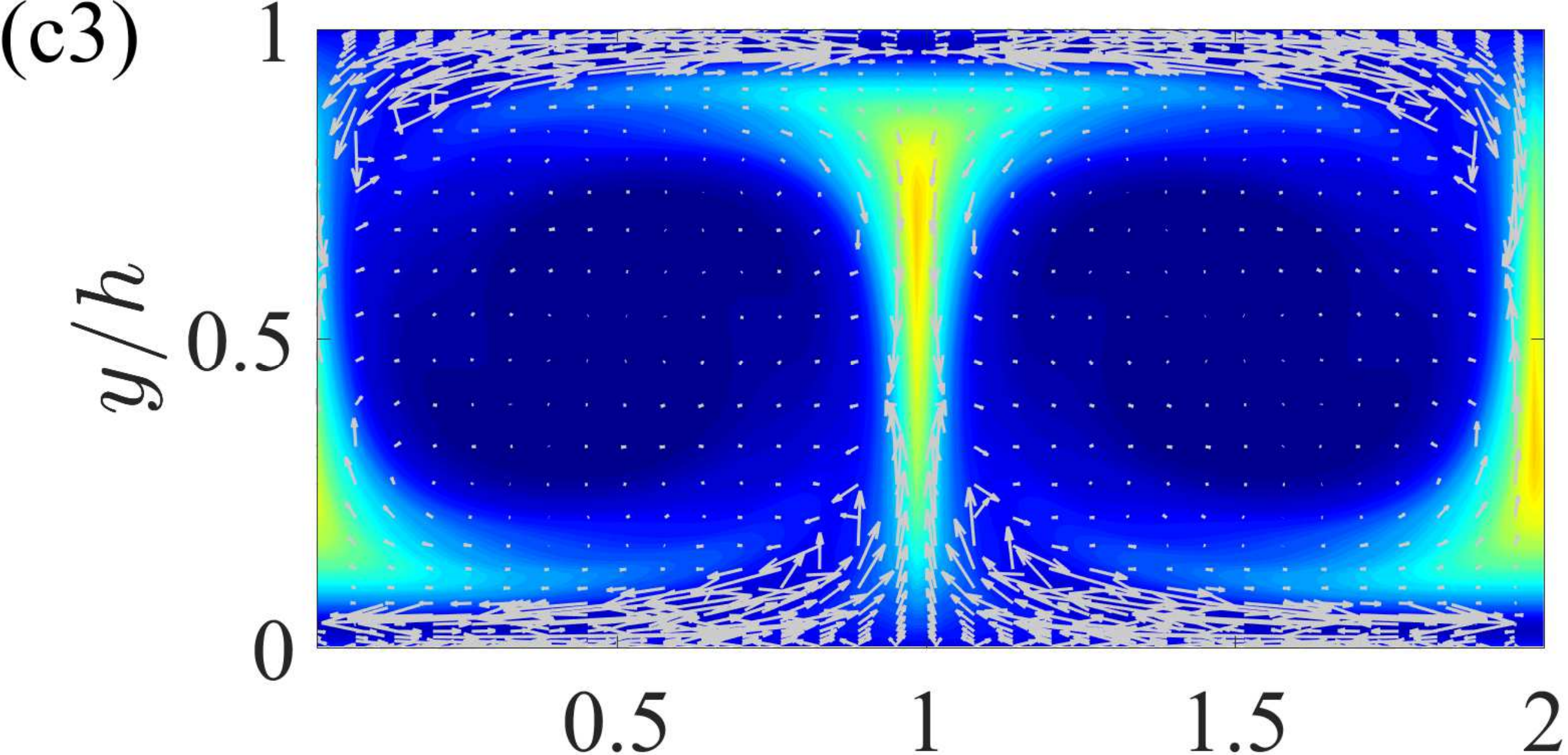}\\
   \includegraphics[width=1\hsize]{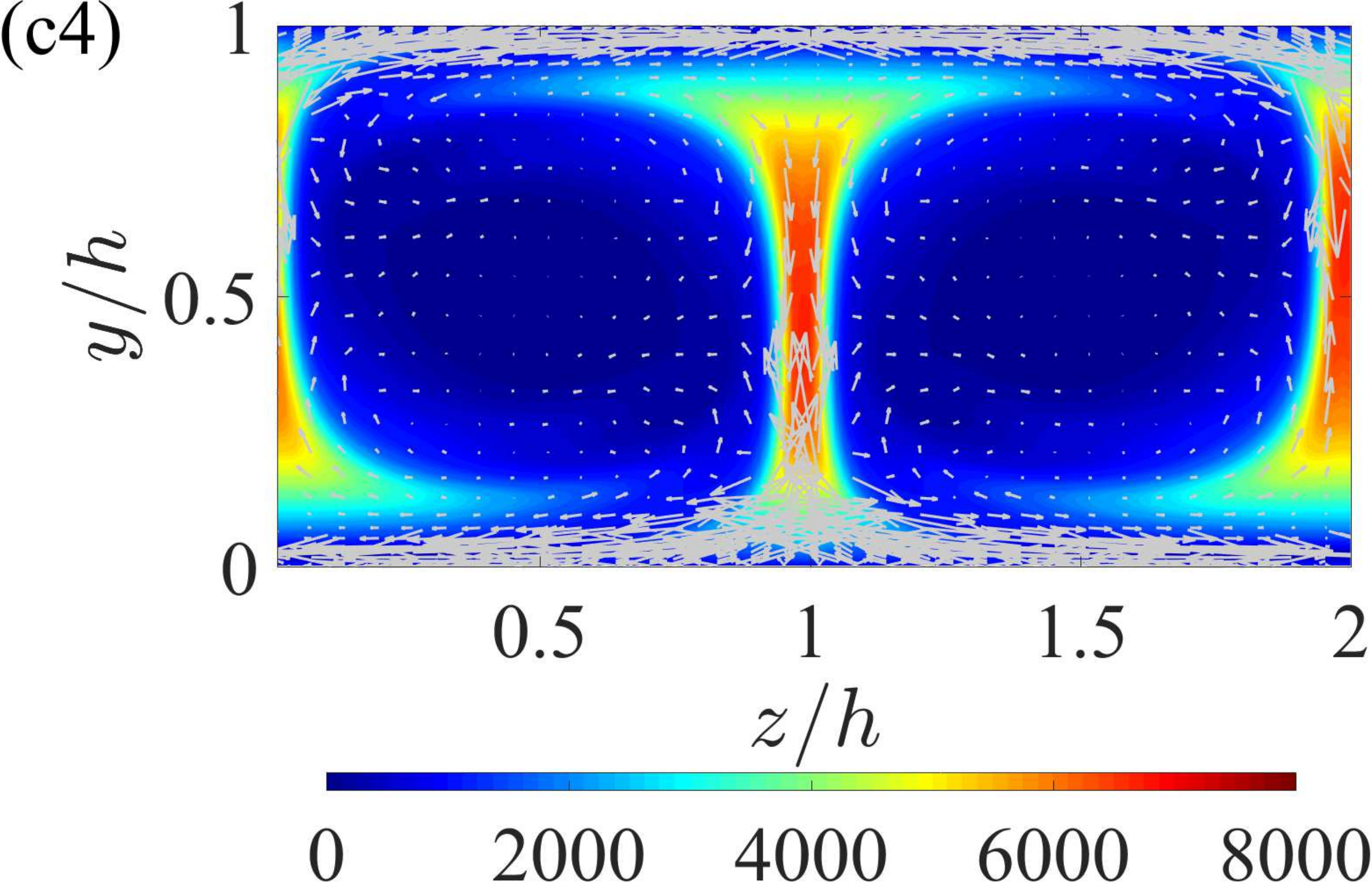}
  \end{minipage}
 \end{tabular}
\caption{Temporal evolution of the (a1--a4) velocity field, (b1--b4) viscosity force field, and (c1--c4) viscoelastic force field on an arbitrary cross-sectional plane for $\Wew=1000$. 
The column of the figure array indicates the time evolution; (a-c1), (a-c2), (a-c3), and (a-c4) present the distributions at $t=t_1$, $t_2$, $t_3$, and $t_4$, respectively, which are defined in Fig.~\ref{fig:stdiag}(b). 
The colors and the arrows in the figures represent the following: (a1--a4) the same things as in Fig.~\ref{fig:rolls}; (b1--b4) the streamwise component and the in-plane components of the viscous force vector, respectively; and (c1--c4) the streamwise normal component of the conformation tensor $c_{xx}$ and the in-plane components of the viscoelastic force vector $(E_y, E_z)$, respectively.}
\label{fig:wi1000}
\end{figure*}

The cross-sectional views of the roll-cell structure with the pulsatile motion are presented in Fig.~\ref{fig:wi1000}. 
Each column of the figure array gives the distributions of (a1--a4) velocities, (b1--b4) viscous forces, and (c1--c4) viscoelastic forces, at $t=t_1$, $t_2$, $t_3$, and $t_4$, from the top to the bottom. 
In panels (a1--a4), the fluctuating streamwise velocity $u^\prime /U_\mathrm{w}$ is shown by the contour, whereas the cross-flow vectors are shown by the black arrows. 
The roll-cell structure was enhanced in the phase from (a1) to (a3) and damped again in (a4). 
In panels (b1--b4), the streamwise viscous force is shown by the contour, whereas the in-plane viscous force vectors are shown by the black arrows. 
Comparing panels (b1--b4) to (a1--a4), one can see that the distribution of the in-plane viscous force was counter rotating against the vortical motion of the roll cells and that the positive/negative peak of the streamwise viscous force corresponded to the negative/positive peak of the fluctuating streamwise velocity, indicating that the viscous force was always counteracting the flow structure. 

\begin{figure}[t]
\begin{center}
	\includegraphics[width=0.5\hsize]{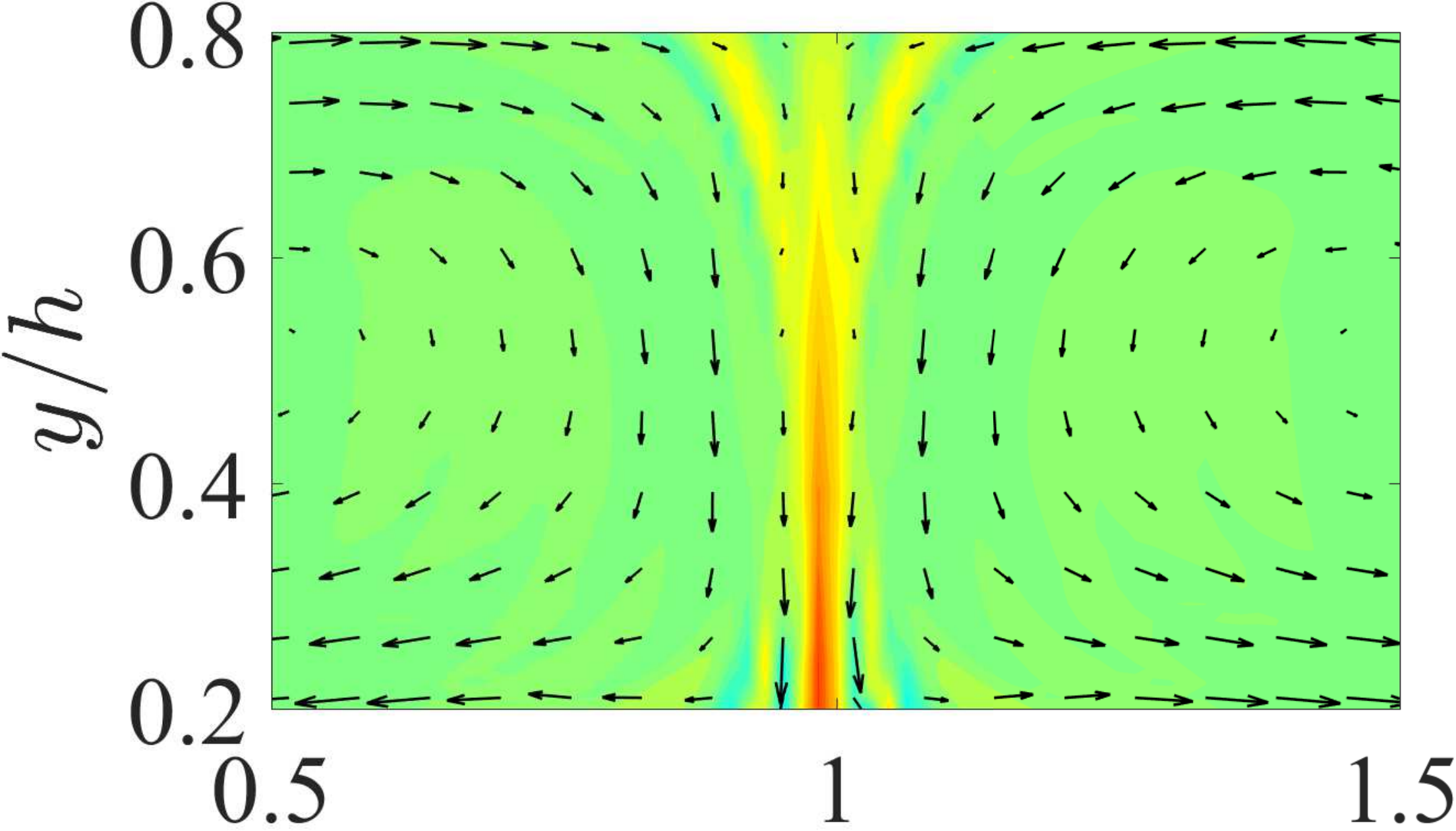}\\
        \vspace{0.5em}
	\includegraphics[width=0.5\hsize]{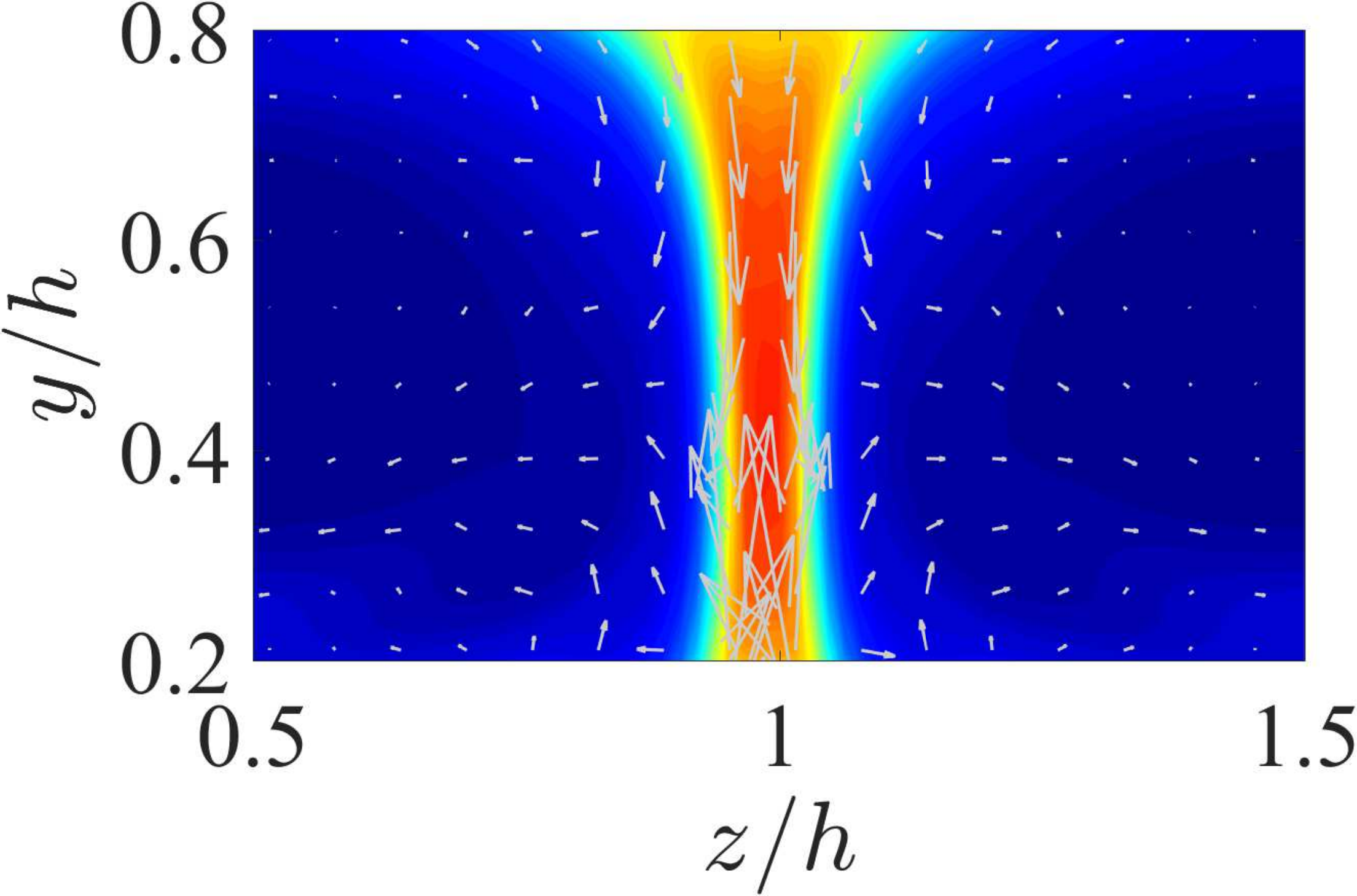}
	\caption{Magnified view of the region around $(z/h, y/h)=(1, 0.5)$ in (top) Fig.~\ref{fig:wi1000}(b4) and (bottom) Fig.~\ref{fig:wi1000}(c4). The contour and the length scale of the white arrows are the same as in the original figures.}
	\label{fig:mag}
\end{center}
\end{figure}

In panels (c1--c4) of Fig.~\ref{fig:wi1000}, the contour shows the streamwise component of the conformation tensor $c_{xx}$, whereas the white arrows represent the in-plane viscoelastic force $(E_y, E_z)$, where 
\begin{eqnarray}
E_y= \frac{1-\beta}{\Wew} \left( \pd{c_{yy}}{y^*} + \pd{c_{yz}}{z^*} \right), \\
E_z= \frac{1-\beta}{\Wew} \left( \pd{c_{zy}}{y^*} + \pd{c_{zz}}{z^*} \right).
\end{eqnarray}
The component $c_{xx}$ was dominant in the trace of $c_{ii}$ in this flow, and the physical meaning can be interpreted as the polymer stretching mainly in the streamwise direction. 
It is shown that $c_{xx}$ was almost zero inside the roll cells, whereas it increased on the edge of the roll cells: see $z/h = 0$ (or 2) and 1 of Fig.~\ref{fig:wi1000}. 
A similar tendency could be observed for the in-plane viscoelastic force distribution. 
In particular, comparing the in-plane viscous force at the same instance, one can see that the magnitude of the viscoelastic force was generally smaller than the viscous force inside the roll cells (the viscous and viscoelastic forces in panels (b1--4) and (c1--4) are shown in the same unit scale length of the arrows), whereas on the edges of the roll cells the viscoelastic force was as large as the viscous force. 
Such tendency in the behavior of the viscoelastic force is more easily seen in Fig.~\ref{fig:mag}, where the region around $(z/h,y/h)=(1,0.5)$ of Fig.~\ref{fig:wi1000}(b4) and (c4) is magnified. 
Inside the vortex, the white arrows of the viscoelastic force (in the bottom figure) are clearly shorter than the black arrows of the viscous force (in the top figure), whereas on the edge of the vortices (around $z/h=1$) the white arrows are longer than the black arrows. 
It should also be noted that, in the region of $0.2 \leq y/h \leq 0.4$ around $z/h=1$, the viscoelastic force was in the same direction as the cross-flow pattern. 
This feature of vortical (or upward ejecting) motion support of the roll-cell structure could not be observed for the viscous force. 

\begin{figure}[t]
\begin{center}
	\includegraphics[width=0.666\hsize]{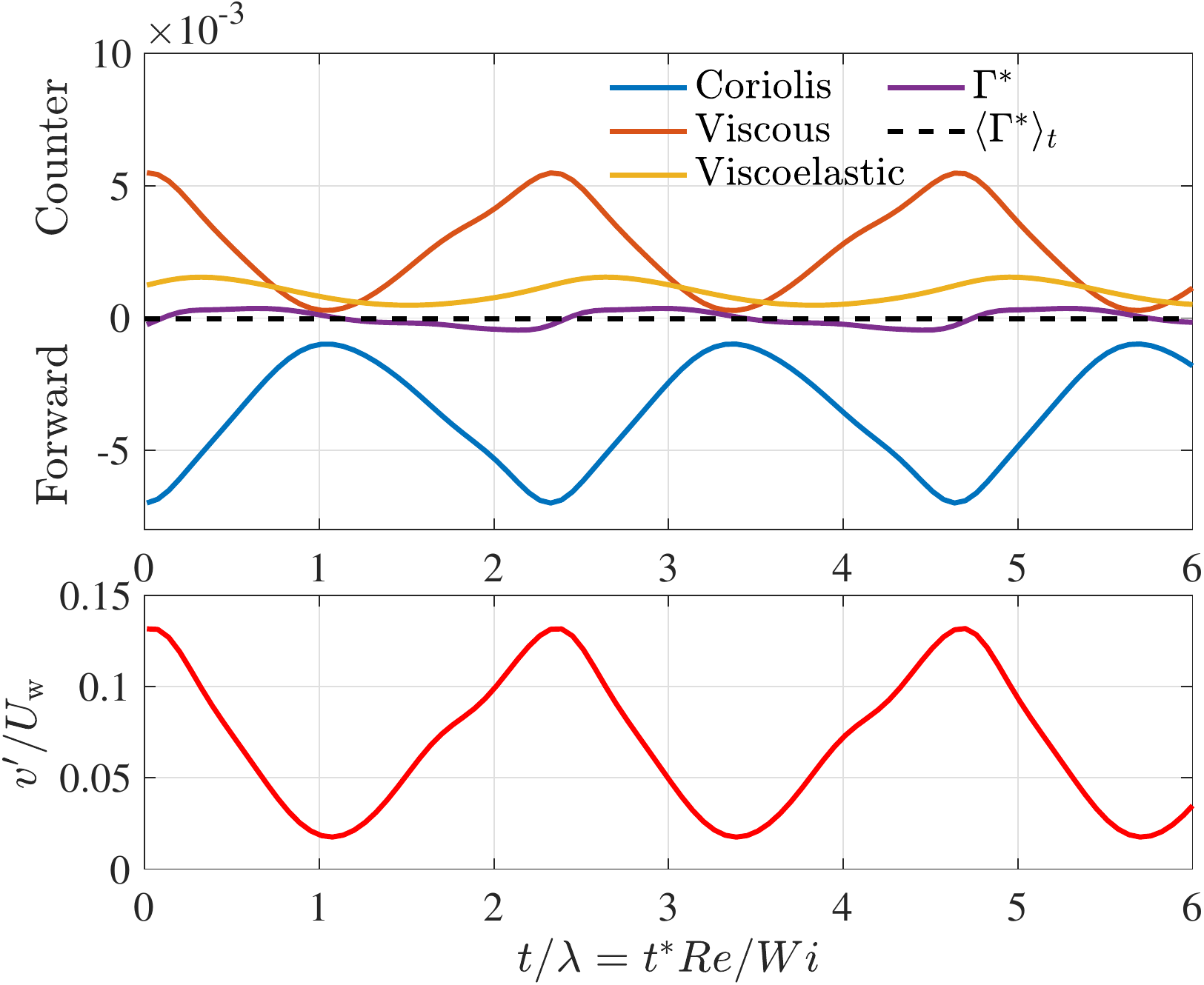}
	\caption{Time sequences of (top) the torques acting on the roll cell located at $(z/h,y/h)=(1.5,0.5)$ and (bottom) the wall-normal velocity $v^\prime/U_\mathrm{w}$ at $(z/h,y/h)=(1,0.5)$ at $\Wew=1000$. The total torque $\Gamma^*$ is the sum of the torque contributions by the Coriolis, viscous, and viscoelastic forces. Its time-averaged value can be confirmed to be approximately zero $\langle \Gamma^* \rangle_t \approx 0$. }
	\label{fig:torque}
\end{center}
\end{figure}

Figure~\ref{fig:torque} shows the time sequence of the torque acting on the roll cell on the right-hand side of Fig.~\ref{fig:wi1000}, i.e., the torque located at $(z/h,y/h)=(1.5,0.5)$ integrated over the right half of the cross section shown in Fig.~\ref{fig:wi1000}, compared with the wall-normal velocity $v^\prime$ at the center of the computational domain $(z/h,y/h)=(1,0.5)$. 
As the roll cell at $(z/h,y/h)=(1.5,0.5)$ was rotating in the clockwise direction, the negative/positive torque in the figure indicates a forward/counter contribution. 
As shown, the roll cell was driven by the forward torque of the Coriolis force, and the viscous and viscoelastic forces were counteracting to the roll cell. 
Although the viscous torque was in the anti-phase of the Coriolis torque, the viscoelastic torque had a certain delay compared with it. 
The temporal change in the wall-normal velocity $v^\prime$, which represents the cell-rotation magnitude, was correlated well with the Coriolis torque. 
It is also seen that there is a small bump in the $v^\prime$ profile during the growth phase, and this might have been due to the viscoelastic contribution. 
The viscoelasticity-induced negative torque on the vortex suppressed the flow structure, in particular, of the streamwise vortex, and DR occurred as a result.

\begin{figure}[t]
\begin{center}
	\includegraphics[width=0.7\hsize]{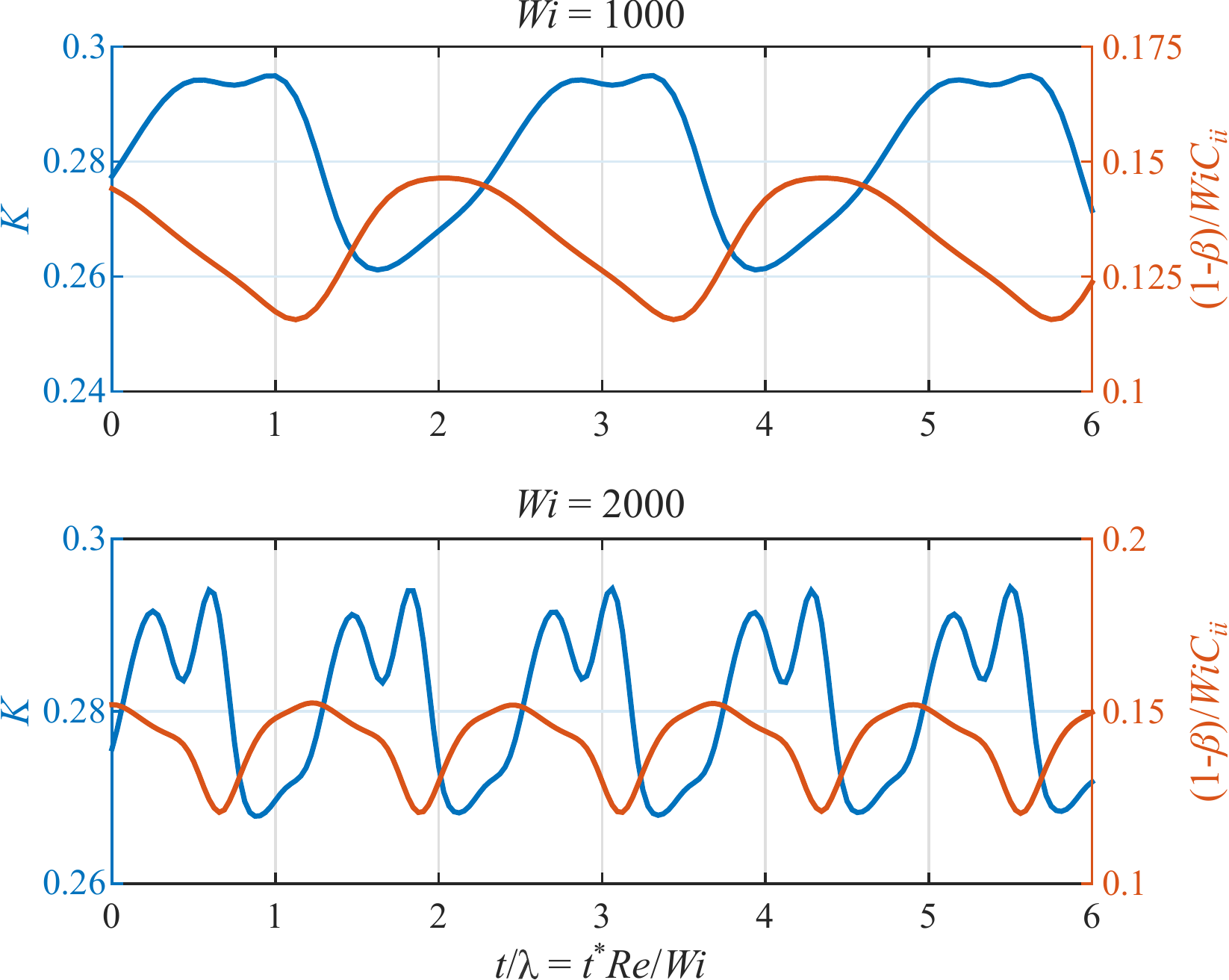}
	\caption{Time sequence of the kinetic energy and the trace of the conformation tensor integrated across the cross section of the channel for (top) $\Wew=1000$ and (bottom) $\Wew=2000$. Blue, $K$ defined by Eq.~(\ref{eq:allK}); orange, $(1-\beta)\Wew {\rm tr}(C_{ii})$.}
	\label{fig:k_c}
\end{center}
\end{figure}

\begin{figure}[t]
\begin{center}
	\includegraphics[width=0.7\hsize]{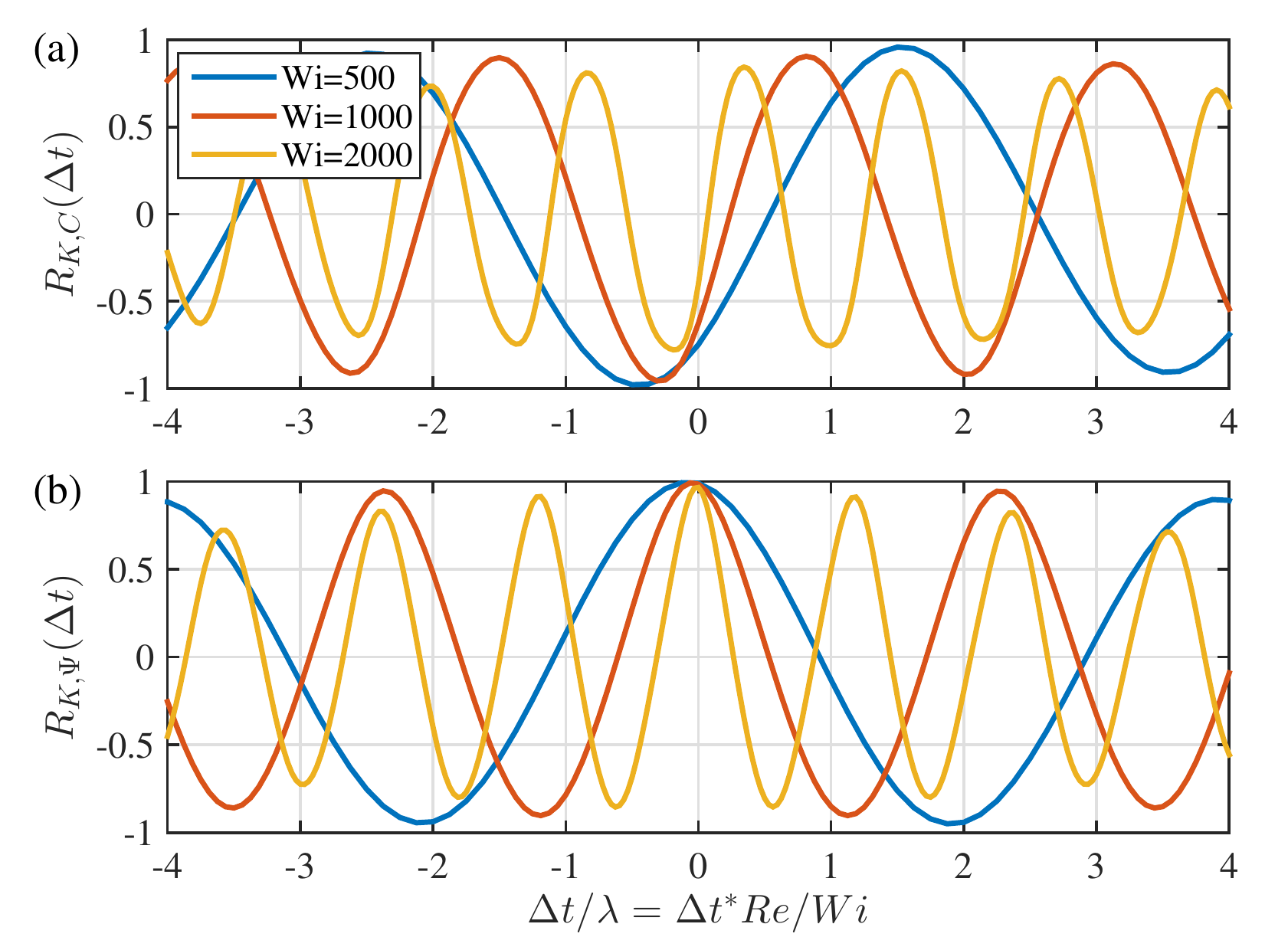}
	\caption{Temporal two-point cross-correlation function (a) between the integrated kinetic energy $K$ and the integrated trace of the conformation tensor $C_{ii}$, and (b) between $K$ and the integrated energy transfer term $\Psi_{ii}$ for three different Weissenberg numbers: blue, $\Wew=500$; red, 1000; and orange, 2000.}
	\label{fig:tcc2}
\end{center}
\end{figure}

The other noteworthy tendency here is the delay in the response of the viscoelastic force to the variation in the flow structure: whereas the roll-cell structure and the viscous force became most significant at $t=t_3$, it was at $t=t_4$ that the viscoelastic force became the largest. 
Such time lag between the variation in the flow field and the viscoelastic tensor is shown in more detail in Fig.~\ref{fig:k_c}, where the time series of the volume-averaged kinetic energy $K$ and the trace of $C_{ij}$, defined below, are compared for the cases of $\Wew=1000$ and 2000:
\begin{eqnarray}
K&=&\frac{1} {L_x L_y L_z}  \int^{L_x}_0 \int^{L_z}_0 \int^{L_y}_0 k \; {\rm d}x{\rm d}y{\rm d}z  =\frac{1} {L_y L_z}  \int^{L_z}_0 \int^{L_y}_0 k \;{\rm d}y{\rm d}z, \label{eq:allK} \\
C_{ii}&=&\frac{1} {L_y L_z}  \int^{L_z}_0 \int^{L_y}_0 (c_{xx}+c_{yy}+c_{zz}) {\rm d}y{\rm d}z \label{eq:C},
\end{eqnarray}
where $k=(u^2+v^2+w^2)/2$. 
In the figure, the time was normalized by the relaxation time $\lambda$ rather than by the shear rate or the turnover time of the system rotation.
Both $K$ and $C_{ii}$ pulsated in time, although a significant delay existed between the temporal variation of $K$ and $C_{ii}$, as seen in Fig.~\ref{fig:k_c}. 
To quantify the time delay between the flow structure and the viscoelastic stress, we evaluated the temporal cross-correlation between the time series of $K$ and $C_{ii}$ , which is shown in Fig.~\ref{fig:tcc2}(a) for three different $\Wew$ cases. 
Note here that the pulsatile motion that occurred transiently before reaching a steady state was examined for $\Wew=500$.
For all the cases, the cross-correlation between $K$ and $C_{ii}$ at $\Delta t = 0$ was negative, implying an antiphase with a certain time lag between the vortical motion and the reacting conformation-tensor field. 
The first positive peak of the cross-correlation indicates the magnitude of the time lag. 
Figure~\ref{fig:tcc2}(a) reveals the first positive peaks at $\Delta t/\lambda = 1.5$, 0.81, and 0.34 for $\Wew=500$, 1000, and 2000, respectively. 
This indicates that the time delay of the viscoelastic stress against the growing vortical motion would be on the order of the relaxation time $\lambda$, although it still could not be scaled well by using $\lambda$ only. 
Moreover, the constant magnitudes of both the positive and the negative peaks imply that $K$ and $C_{ii}$ pulsated similarly with the same frequency and constant time lag.
Table~\ref{tab:time} summarizes the ratio between the relevant time scales, where one can see that $t_{\rm P}$ could be scaled by $t_{\rm cell}$ rather than by the relaxation time $\lambda$.
This implies that the increased $t_{\rm cell}$ owing to an elasticity-induced negative torque against the cell rotation was always comparable to the pulsation period.
Although both $t_{\rm P}$ and $t_{\rm lag}$ could be comparable to $\lambda$, their ratios clearly depended on $\Wew$.
As for $t_{\rm lag}$, the scaling by $t_{\rm P}$ rather than by $t_{\rm cell}$ seemed reasonable.
These scalings suggest that the pulsation period was related to the flow dynamics as well as the rheological characteristics, despite the viscoelastic instability.
This conclusion should be examined further at different values of ${\Rew}$ and $\Omega$.

\begin{table}
\begin{center}
\caption{Time scale ratios relevant to the pulsation period $t_{\rm P}$ or the $K$--$C_{ii}$ time lag $t_{\rm lag}$: $t_{\rm cell}=2\pi \delta/\max ( \langle v'\rangle_t)$ is the turnover time of the cell rotation, whereas $t_\Omega=2\pi/\Omega_z$ is that of the system rotation.}
\label{tab:time}
 \begin{tabular}{rrcc ccc ccc}
\hline\hline
  $\Wew$
& ${\rm De}$
& $t_{\rm cell}/t_\Omega$
& $\lambda/t_{\rm cell}$
& $t_{\rm P}/\lambda$
& $t_{\rm P}/t_{\rm cell}$
& $t_{\rm lag}/\lambda$
& $t_{\rm lag}/t_{\rm cell}$
& $t_{\rm lag}/t_{\rm P}$ \\\hline
Newt.   &--     &1.30   &--     &--     &--     &--     &--     &--     \\
10	&0.4	&1.22	&0.010 	&--	&--	&--	&--	&--	\\
100	&4	&1.31	&0.097 	&--	&--	&--	&--	&--	\\
500	&20	&2.26	&0.282 	&3.95	&1.11	&1.50	&0.423	&0.380	\\
1000	&40	&2.81	&0.453 	&2.32	&1.05	&0.81	&0.367	&0.349	\\
2000	&80	&3.10	&0.822 	&1.16	&0.95	&0.34	&0.280	&0.294	\\
\hline\hline
 \end{tabular}
\end{center}
\end{table}

\begin{figure}[t]
\begin{center}
	\includegraphics[width=0.7\hsize]{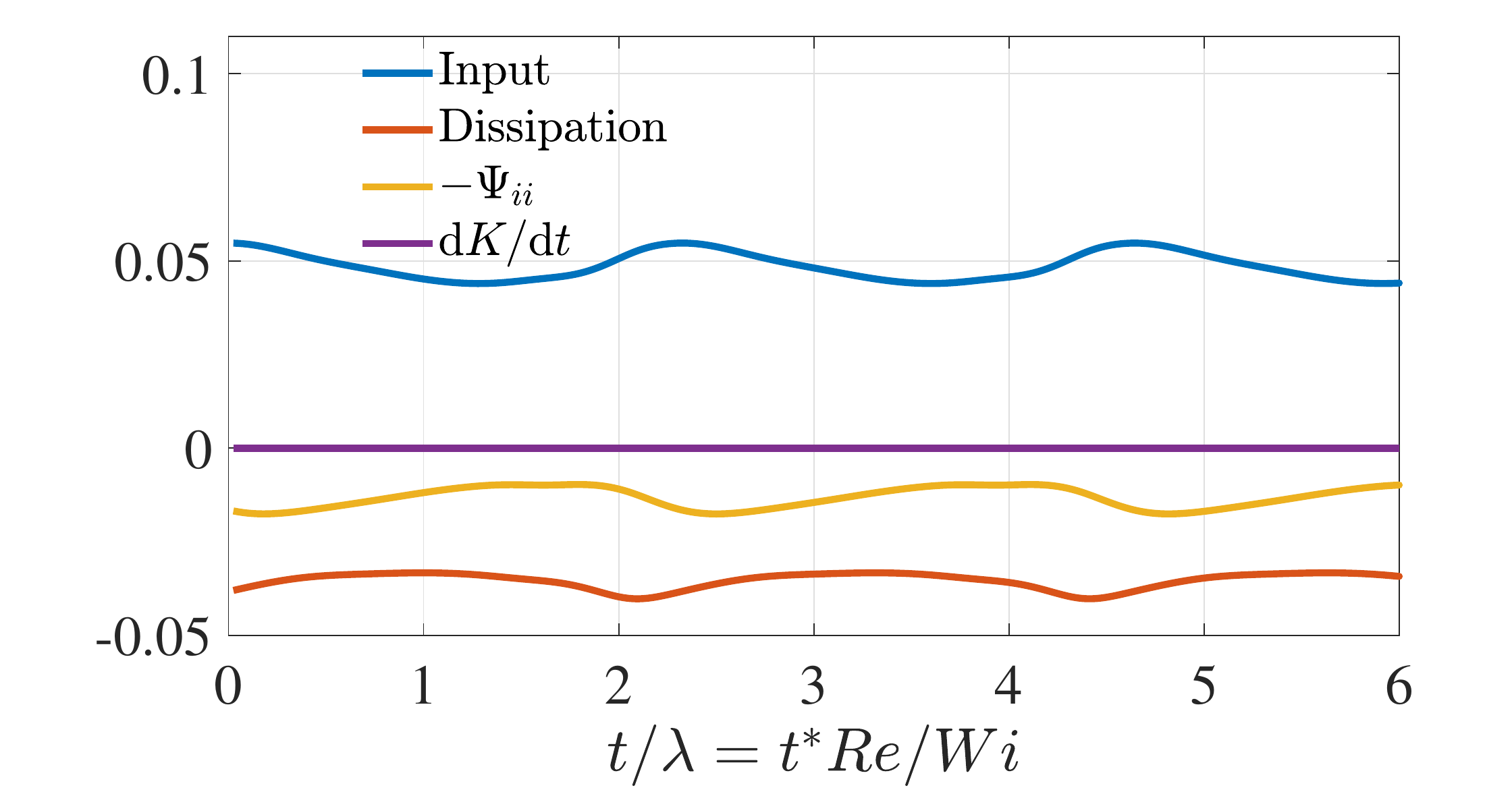}
	\caption{Time sequence of the energy input, the dissipation, and the energy exchange term $\Psi_{ii}$ in Eq.~(\ref{eq:K}) at $\Wew=1000$.}
	\label{fig:tcc}
\end{center}
\end{figure}

To better understand the mechanism that sustains the periodic behavior in the high $\Wew$ cases, we examined the energy transfer between the kinetic energy of the flow field and the `elastic' energy of the polymer. 
The transport equation of the (non-volume-integrated) kinetic energy $k$ is written as
\begin{eqnarray}
\frac{{\rm D} k}{{\rm D} t} &=& \pd{u_i p}{x_i}  + \frac{\beta}{\Rew} \nabla^2 k - \frac{\beta}{\Rew} \left( \pd{u_i}{x_j} \right)^2 \nonumber \\
&&+ \frac{1-\beta}{\Wew} \left( \pd{u_i c_{ij}}{x_j} - c_{ij} \pd{u_i}{x_j} \right). 
\label{eq:dk}
\end{eqnarray}
The last term $c_{ij} \partial u_i / \partial x_j$ in the right-hand side appears also in Eq.~(\ref{eq:Giesekus}) with the opposite sign, which indicates that this term physically represents the energy transfer between $k$ and $c_{ii}$. 
The contribution of the Coriolis force vanished because it was always normal to the velocity vector and was cancelled out in the $k$-transport equation. 
Taking the volume average of Eq.~(\ref{eq:dk}) over the domain computed, one obtains a global energy balance of $K$:
\begin{eqnarray}
\frac{{\rm d} K}{{\rm d}t} &=& \frac{\beta}{\Rew} \frac{1}{L_x L_z} \int^{L_x}_0 \int^{L_z}_0 \left(- \left. \pd{k}{y} \right|_{y=0} - \left. \pd{k}{y}  \right|_{y=h} \right) {\rm d}x {\rm d}z \nonumber \\
&& -\frac{\beta}{\Rew} \frac{1}{V} \iiint \left( \pd{u_i}{x_j} \right)^2 {\rm d}V - \Psi_{ii}, 
\label{eq:K}
\end{eqnarray}
\noindent
where $V^{-1} \iiint \; {\rm d}V$ represents taking the volumetric average over the computational domain and $\Psi_{ii} = V^{-1} \iiint c_{ij} \partial u_i / \partial x_j {\rm d}V$ is the volume-averaged energy transfer term between the flow and the additive. 
Note here that $K$ represents the total kinetic energy including the mean flow, whereas $K^\prime$ shown in Fig.~\ref{fig:K_t} consists only of the fluctuating component $u_i^\prime$. 
In the right-hand side of Eq.~(\ref{eq:K}), the first term corresponds to the energy input to the system as the work to maintain the wall velocity $U_{\rm w}$.
Without viscoelasticity, the kinetic energy should be maintained by the balance between this source term and the viscous dissipation term (the second term). 
The energy transfer term $\Psi_{ii}$ comes into play in the case of the viscoelastic fluid. 
Figure~\ref{fig:tcc2}(b) shows the temporal cross-correlation function between $K$ and $\Psi_{ii}$. 
The cross-correlation was almost unity at $\Delta t$ = 0, which means that there was no time delay between the temporal variations of $K$ and $\Psi_{ii}$. 
If the kinetic energy of the flow structure increased, the energy was forthwith transferred to the additive through $\Psi_{ii}$. 
Figure~\ref{fig:tcc} shows the temporal variations of the terms in Eq.~(\ref{eq:K}) at $\Wew=1000$. 
All the terms had the same periodicity, although there was some phase shifting between them. The energy exchange term $-\Psi_{ii}$ had a certain time delay against the input and kept the negative contribution in the global energy balance, which indicates that the energy was generally transferred from the flow to the additive. It is also shown that the fluctuation of the unsteady term is very small compared to the other terms, which means that the input, dissipation, and energy exchange terms are nearly in equilibrium and the pulsatile motion is caused by small imbalance between them. 
More detailed analysis is necessary to elucidate the energy balance between the mean flow, the roll cell, and the polymers, and its relation to the characteristic features of EIT. 

\section{Conclusion}
\label{conclusion}

We numerically investigated the flow structure of the laminar rotating plane Couette flow of a Giesekus viscoelastic fluid by means of direct numerical simulations (DNSs).
The laminar case of $\Rew=25$ and $\Omega=10$ was focused on with a particular interest on how a two-dimensional (2D) steady roll-cell structure that appears in the Newtonian fluid case is modulated by the fluid viscoelasticity. 
The addition of the viscoelasticity was not found to change the spatial structure of the 2D straight roll cells, but was found to affect the temporal behavior of the structure. 

We demonstrated that the viscoelasticity increased the growth rate of the velocity disturbance to the roll cell. 
At high enough $\Wew$, the viscoelasticity gave rise to a pulsatile flow state, in which the spatial structure of the roll cell was unchanged but its magnitude was periodically enhanced and damped in time, keeping its spatial homogeneity in the streamwise direction. 
Such an additional instability due to the viscoelasticity was observed as the additive relaxation time $\lambda$ became comparable to the turnover time of the cell rotation $t_{\rm cell}$. 
Independently of $\Wew$, the pulsation period $t_{\rm P}$ was approximately equal to $t_{\rm cell}$. 
In the pulsatile case, the mean magnitude of the cell rotation was suppressed ($t_{\rm cell}$ decreased as a result) and the skin friction was decreased from the steady-state case. 
There was a certain time lag in the periodic variations in the kinetic energy and the additive energy. 
The time lag $t_{\rm lag}$ was also on the order of $\lambda$, although the scaling with $t_{\rm cell}$ or $t_{\rm P}$ was likely, e.g., $t_{\rm lag}/t_{\rm P} = 0.3$--0.4.
In the current flows, the global energy transfer was generally from the flow to the additive; however, this energy flux was also pulsative with a time scale of the relaxation. 
In addition to this net energy path, the finding that the viscoelasticity induced a negative torque on the streamwise vortex (even without a turbulent background) was linked to the DR mechanism, whereas the route to EIT was found as the onset of unsteadiness in the form of temporal pulsation without spatial variation in the roll cells.

The present work sheds light on the viscoelasticity effect of destabilizing the longitudinal vortex in a canonical wall-bounded flow.
The above conclusions have been drawn from a limited case with a single set of $\Rew$, $\Omega$, $\alpha$, and $\beta$. 
For instance, at a higher $\Rew$ of 100, \cite{Nimura18} reported that wavy 3D roll cells were stabilized and straightened into 2D roll cells with their spanwise flow motions being weakened, whereas the streamwise streaks were maintained.
This viscoelasticity effect is closely related to polymer drag reduction, although it is different quantitatively in that it triggers the presently observed pulsatile flow.

Further parametric DNS study is necessary for wider ranges of parameters, which would require a much larger computational domain and time.

\section*{Acknowledgment}

This work was partially supported by the Grant-in-Aid for Young Scientists (A) (16H06066) from the Japan Society for the Promotion of Science (JSPS).
T.K. was supported by a Research Fellowship for Young Scientists (17J04115) from the JSPS. 
The present simulations were performed on an SX-ACE supercomputer (NEC) both at the Cybermedia Centre of Osaka University and at the Cyberscience Centre of Tohoku University. 
We also thank the reviewers for their insightful and constructive comments. 

\section*{Appendix A. Grid-convergence and -robustness tests}

\begin{figure*}[t]
\begin{center}
	\includegraphics[width=0.7\hsize]{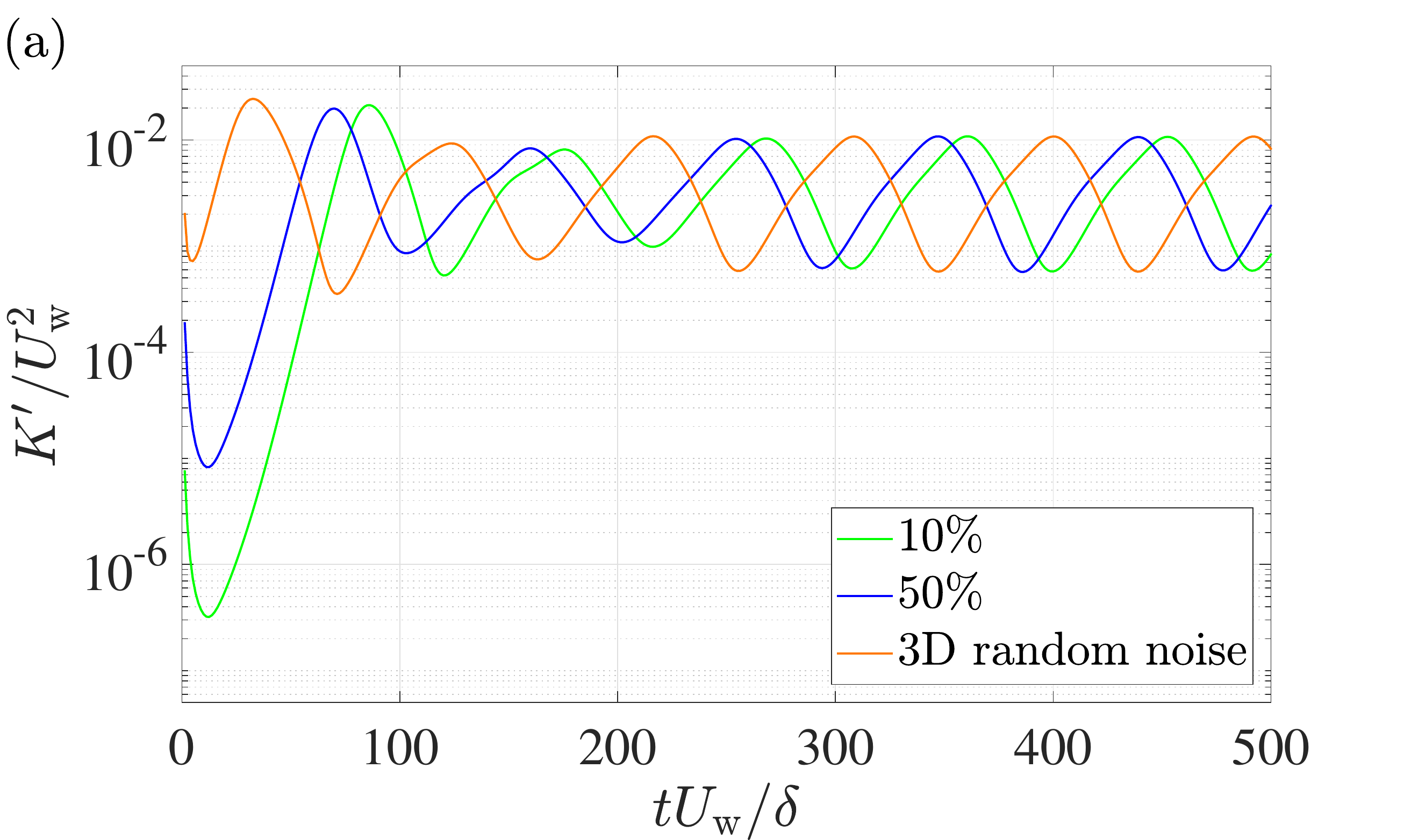}\\
	\includegraphics[width=0.7\hsize]{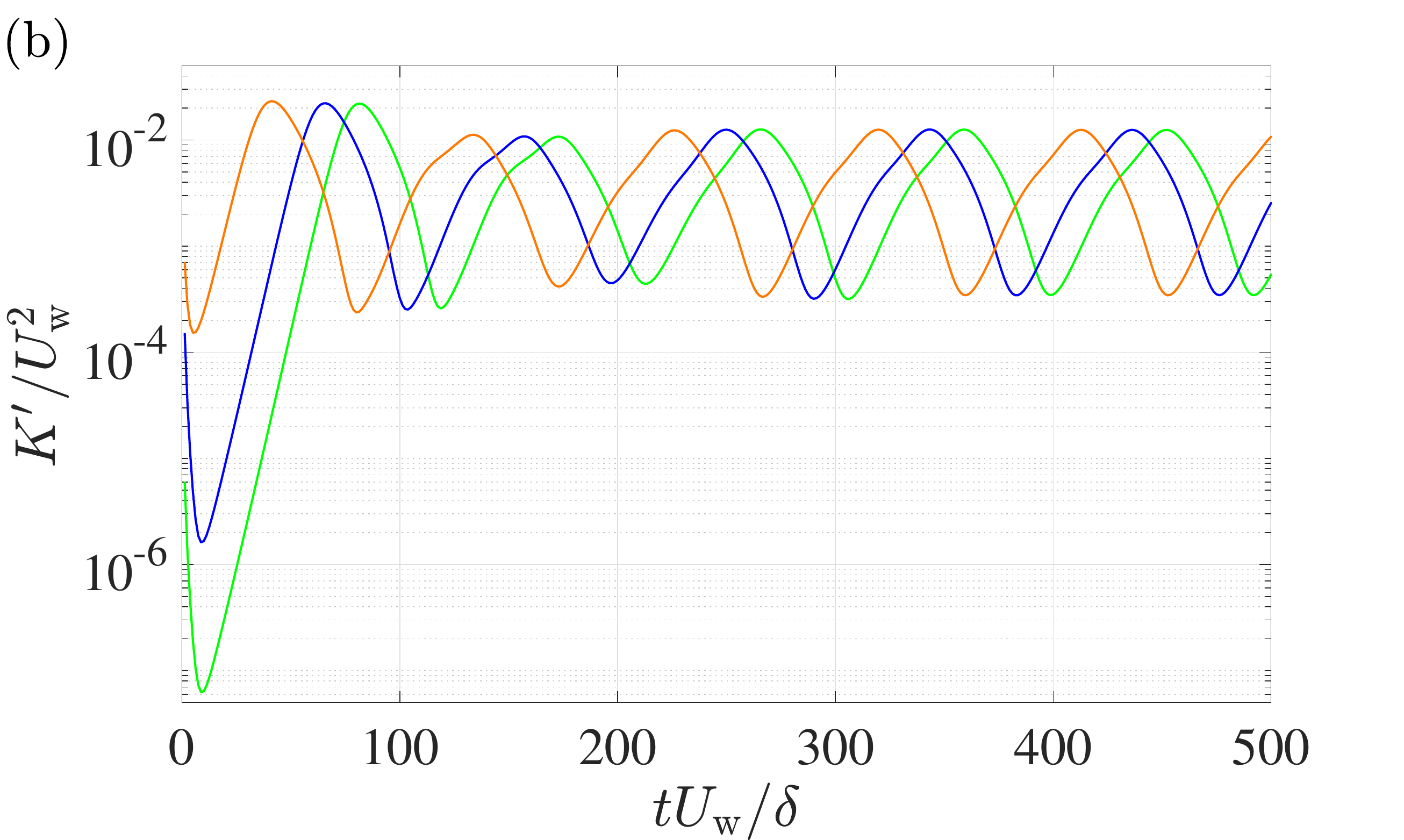}
	\caption{Time evolution of the volume-averaged kinetic energy $K$ for $\Wew=1000$ and its dependencies on the number of grid points and the initial disturbance of the DNS, where various 3D noise was given as the disturbance. A grid-convergence test was done by comparing between (a) $64^3$ and (b) $128^3$ grid points. As for the initial disturbance, 3D random noise with a magnitude of either $0.1U_{\rm w}$ or  $0.5U_{\rm w}$ and a fully developed turbulence were tested.} 
	\label{fig:cnvt}
\end{center}
\end{figure*}

We conducted a convergence test with respect to the grid resolution by changing the grid numbers as $(N_x, N_y, N_z) = (128, 128, 128)$ down to $(64, 64, 64)$. In addition, we also examined the robustness of the present results, in particular, the pulsatile motion at $\Wew = 1000$ for rather strong or three-dimensional (3D) noise. 
We prepared two fields of 3D random noise as an initial random disturbance to all velocity components, but with two different levels of magnitude: one had an intensity of, at most, 10\% of the wall speed $U_{\rm w}$, and the other case was comparable to half of $U_{\rm w}$ (50\%). 
As a more realistically disturbed flow, the fully developed turbulence at $\Rew = 500$ was also tested to demonstrate the onset of the pulsatile motion. 
Figure~\ref{fig:cnvt} shows all six test cases with different numbers of grid points and different types of initial disturbances. 
As similarly seen in the case of weak disturbance (Fig.~\ref{fig:K_t}), after an exponential growth and/or transient overshoot, well-defined periodic oscillations can be confirmed in Fig.~\ref{fig:cnvt}.
Even with strong disturbances, the initial turbulent motions were rapidly attenuated, and only the roll cells robustly survived and exhibited the pulsatile motion with identical periodicity. 
Essentially the same results were obtained in Fig.~\ref{fig:cnvt}(a) and (b), implying an adequate spatial resolution.

\end{document}